\documentclass[useAMS,preprint]{aastex} 

\usepackage{amsmath}
\usepackage{amsbsy}

\input epsf
\usepackage{amsmath,amssymb,subfigure}
\usepackage{graphicx}
\usepackage{epsfig}
\usepackage{color}
\usepackage{hyperref}
\usepackage{algorithmicx}
\usepackage{algorithm}
\usepackage{algpseudocode}

\hypersetup{
    colorlinks=true,
    linkcolor=red,
    citecolor=blue,
}

\newcommand{\be}{\begin{equation}}
\newcommand{\ee}{\end{equation}}
\newcommand{\beq}{\begin{equation}}
\newcommand{\eeq}{\end{equation}}
\newcommand{\beqa}{\begin{eqnarray}}
\newcommand{\eeqa}{\end{eqnarray}}

\newcommand{\ls}{\mathrel{\raise0.27ex\hbox{$<$}\kern-0.70em \lower0.71ex\hbox{{
$\scriptstyle \sim$}}}}

\newcommand{\clim}{\chi^2_{\rm lim}}
\newcommand{\cmin}{\chi^2_{\rm min}}
\newcommand{\dchi}{\Delta\chi^2}

\newcommand{\ourCode}{DALE$\chi$ }

\newcommand{\acronym}{(Direct Analysis of Limits via the Exterior of $\chi^2$) }
\newcommand{\repo}{\url{https://github.com/danielsf/Dalex.git} } 

\begin{document} 

\title
{Accelerated Parameter Estimation with \ourCode}
\author
{Scott F.\ Daniel$^1$ and Eric V.\ Linder$^{2,3}$\\
$^1$Department of Astronomy, University of Washington, Seattle, WA\\ 
$^2$Berkeley Center for Cosmological Physics \& Berkeley Lab, 
University of California, Berkeley, CA 94720\\ 
$^3$Energetic Cosmos Laboratory, Nazarbayev University, Astana, 
Kazakhstan 010000}



\date{\today}


\begin{abstract} 
We consider methods for improving the estimation of constraints on a
high-dimensional parameter space with a computationally expensive
likelihood function.   In such cases Markov chain Monte Carlo (MCMC)
can take a long time to converge and concentrates on finding the maxima
rather than the often-desired confidence contours for accurate error
estimation.  We employ \ourCode \acronym for determining confidence
contours by minimizing a cost function parametrized to incentivize
points in parameter space which are both on the confidence limit and
far from previously sampled points.  We compare \ourCode to the
nested sampling algorithm implemented in MultiNest on a toy likelihood
function that is highly non-Gaussian and non-linear in the mapping between
parameter values and $\chi^2$.  We find that in high-dimensional cases 
\ourCode finds the same confidence limit as MultiNest using roughly
an order of magnitude fewer evaluations of the likelihood function.
\ourCode is open-source and available at \repo.
\end{abstract}

\section{Introduction}
\label{sec:intro}

When confronting a model with data, determining 
the best fit parameters is only one
element of the result. For informative comparison one must have a robust
estimation of the confidence intervals for the parameters, and indeed the
full multidimensional confidence contours.
This will become increasingly important with future big data experiments 
where as we strive to achieve precision and accurate cosmology we 
1) need to employ numerous foreground, calibration, or other systematics 
parameters, raising the dimensionality of the fitting space, and 
2) combine data from multiple probes and experiments, requiring 
knowledge of the full multidimensional posterior probability distribution. 

Thus, one wants to {\it optimize\/}
the likelihood function on some parameter space and also
{\it characterize\/} the behavior of the likelihood function in the parameter
space region surrounding that optimum.  For a frequentist confidence
limit (see the appendix of \citet{Daniel:2014}), this means we wish to find
all combinations of parameter values that result in $\chi^2\le\clim$ where
$\clim$ is set by theoretical considerations (e.g.\ a certain $\dchi$ above 
$\cmin$).  For a Bayesian credible limit, we wish to find 
the region of parameter space that maximizes the posterior probability
distribution and then integrate that distribution over parameter space until 
we have a contour that contains some percentage of the total posterior
probability.

Numerous techniques exist for carrying out this process.  The most direct is 
evaluating the likelihood throughout the parameter space, but a grid-based 
approach rapidly becomes intractable as the dimension of the space increases.  
Sampling methods select various points at which to carry out an evaluation, 
and move through parameter space according to some algorithm.  The most common 
such technique in cosmological applications is Markov chain Monte Carlo (MCMC).
This works moderately well (though increasingly slowly) as the dimensionality 
increases but focuses on finding 
the maximum of the posterior probability; the confidence contours are a 
byproduct estimated by 
integrating the distribution of sampled points over the parameter space.

Often we are most interested in the joint confidence intervals for the parameters,
and MCMC can sometimes be slow at converging on these since the algorithm
is more interested in the interior of the contour rather than its boundary.
Moreover, MCMC slows down significantly when confronted with highly
degenerate likelihood surfaces.  Other sampling techniques such as
nested sampling, with  MultiNest \citep{Feroz:2008,Feroz:2009,13062144}
an implementation popular for cosmological applications, can
alleviate those issues, but still do not focus on the confidence interval.

One alternative to these sampling algorithms is Active Parameter Search.
Active Parameter Search (APS) is designed specifically to go after confidence 
intervals \citep{Bryan:2007, brentsthesis, Daniel:2014}.  APS uses a combination
of function optimization and Gaussian Processes \citep{gp} to specifically sample
points on the $\chi^2=\clim$ confidence limit.  In this work, we adopt the
philosophy of APS (targeted sampling of $\chi^2=\clim$ points), but increase
its robustness and speed of convergence by several improvements and new, 
focused search strategies. These are implemented 
using a Nelder-Mead simplex \citep{simplex}
to minimize a cost function of the form
\begin{eqnarray}
F(\vec{\theta})&=&\chi^2(\vec{\theta}) - N(\vec{\theta})\times
E(\chi^2)\times\left(\clim-\cmin\right)\label{eqn:cost}\\
\nonumber\\
E(\chi^2(\vec{\theta}))&=&\begin{cases}\exp\left[\left(\clim-\chi^2(\vec{\theta})\right)/\ell\right],&\text{if }\chi^2(\vec{\theta})>\chi^2_\text{lim}\\
                      1, &\text{otherwise}
                      \end{cases}\nonumber
\end{eqnarray}
where $\vec{\theta}$ is the position in parameter space,
$N(\vec{\theta})$ is the neighbor function, and $E(\chi^2(\vec{\theta}))$ 
is the exterior function. The parameter $\ell$ controls how much optimizations
of $F$ are allowed to explore outside of $\chi^2\le\clim$, and 
$N(\vec{\theta})$ is given by the harmonic mean of the parameter space
distances between $\vec{\theta}$ and the previously-sampled
points such that $\chi^2\le\clim$.  
Minimizing this function drives the Nelder-Mead
simplex to discover points that are both within the confidence contour and far from
previously sampled points, increasing the likelihood that our algorithm will
discover previously unknown regions tracing the true confidence contour.
We refer to this algorithm as \ourCode \acronym.
We find that in problems with complicated, high-dimension likelihood
surfaces, \ourCode achieves convergent, accurate parameter
estimation more rapidly than MCMC and MultiNest.

In Sec.~\ref{sec:methods} we briefly review the methods employed by MCMC,
MultiNest, and conventional APS.  We introduce and discuss the 
strategies of \ourCode in
Sec.~\ref{sec:algorithm}, with their motivations and implementations.
Section~\ref{sec:cartoon} introduces the cartoon likelihood function, 
incorporating aspects of high-dimensionality, non-Gaussianity, and 
degeneracy, on which we
test \ourCode.
Section~\ref{sec:compare} carries out the comparison of parameter estimation
between the techniques.  We conclude in Sec.~\ref{sec:concl}.

Open-source C\texttt{++} code implementing \ourCode is available at the GitHub
repository \repo.  This software depends on 
the publicly available software libraries BLAS, LAPACK (available from
\url{www.netlib.org}), and ARPACK (available from
\url{www.caam.rice.edu/software/ARPACK}).
\ourCode also relies on the KD-Tree algorithm described by \citet{kdtree},
which we use to store the history of parameter-space points explored by
\ourCode,
and the random number generator described by \citet{random}.
Those interested in using \ourCode for their own research
should not hesitate to either contact the authors at \verb|danielsf@astro.washington.edu|
or open an issue on the GitHub repository should they have any questions.

\section{Parameter Estimation Methods} \label{sec:methods}  

\subsection{Markov Chain Monte Carlo} 
\label{sec:mcmc}

MCMC uses chains of samples to evaluate the posterior probability of
the data having a good fit to a point in parameter space, calculated through
a likelihood function and priors on the parameter distribution.  In the
commonly used Metropolis-Hastings algorithm, a point is evaluated to check
if it has a higher probability than the previous point, and accepted or
rejected according to a probability criterion.  The Markov chain part of
the algorithm indicates that only the preceding point affects the next
point, while the Monte Carlo part denotes the random selection of candidate
points and the integration of the probability over the parameter space to
determine the confidence contours. Note that MCMC per se is only trying to
find the maximum; points defining the confidence contours are in some
sense incidental.

\ourCode as discussed below is designed to guide the selection of
desirable points by using all the information already collected, rather than
only the previous point, taking a more global view, and also to accelerate
the robust estimation of the confidence intervals by focusing on points 
informing this result and tracing contours rather than
concentrating points near the probability maximum.

\subsection{Nested Sampling (MultiNest)}
\label{sec:multinest}

Nested sampling \citep{skilling} is a Monte Carlo technique for sweeping 
through the parameter space by comparing a new point with the likelihood 
of all previous points, and accepting it if its likelihood is higher than 
some of them (the lowest of which is then abandoned).  Thus the sample 
moves through nested shells of probability.  MultiNest 
breaks the volume of sampling points 
into ellipsoidal subvolumes and carries out the nested sampling within 
their union, but samples within one ellipsoid at a time.  This concentrates 
the evaluation in a more likely viable region of parameter space while 
allowing for multi-modal distributions. 

MultiNest exhibits accelerated convergence relative to MCMC, and better 
treatment of multimodal probabilities.  However, MultiNest is still designed
to integrate the entire Bayesian posterior over all of parameter space.
Credible limit contours on this posterior are derived as a side-effect,
rather than being intentionally mapped out in detail.  One result of this,
which we present in Section~\ref{sec:sampling}, is that MultiNest 
spends considerable amount of time evaluating $\chi^2$ at points in
parameter space well outside of the credible limit contour.  Such
evaluations are important when trying to
find the full Bayesian evidence of a model, but are not very helpful if
all one is interested in are the multi-dimensional credible limit
contours.  We designed \ourCode for those situations where we want to
accurately map out the confidence intervals of parameters, rather than
characterize the full Bayesian posterior across all of parameter space.

\subsection{Active Parameter Search} 
\label{sec:aps} 

MCMC and MultiNest are both sampling algorithms. 
They attempt to find Bayesian
credible limits by drawing random samples from parameter space in such a way
that the distribution of the samples accurately characterizes the Bayesian
posterior probability distribution on the parameter space.
This strategy directly carries out the {\it optimize\/} step described in the
introduction -- drawing samples with a frequency determined by their posterior
Bayesian probability ensures that the maximum likelihood point in parameter
space has a high probability of being found -- and only incidentally carries out
the {\it characterize\/} step.

The APS algorithm \citep{brentsthesis,Bryan:2007,Daniel:2014}, was designed to 
separately {\it optimize} and {\it characterize} the $\chi^2$ function on 
parameter space.  Separate searches are implemented to find local minima in 
$\chi^2$ -- this is the {\it optimization} -- and then to locate points about 
those minima such that $\chi^2=\clim$, where $\clim$ 
defines the desired frequentist confidence limit (for example in two 
dimensions the 95\% CL is $\clim=\chi^2_\text{min}+6.0$). 
\citet{Daniel:2014} present a way of translating these into Bayesian 
credible limits (see their Section 2.4). These latter searches represent the 
{\it characterization}.  

Though APS improved over MCMC in accurately characterizing complex probability 
surfaces, and its convergence properties were comparable, it remains 
desirable to speed up parameter estimation.  Indeed, as mentioned in the 
previous subsection, nested sampling techniques such as 
MultiNest can also deal well with multimodal distributions and run faster 
than MCMC.  Motivated by the desire for speed, while retaining and 
improving APS' ability 
to efficiently characterize confidence intervals, even for complex 
probability surfaces, we present an algorithm (\ourCode)
with accelerated convergence in mapping the likelihood.

\section{\ourCode \acronym} \label{sec:algorithm} 

Like APS, \ourCode operates by trying to find points directly on the
$\chi^2=\clim$ contour, rather than by inferring that contour from samples
drawn from the posterior.  This requires accurate knowledge of the value of
$\clim$,which, as discussed in Section \ref{sec:limits} below, requires
accurate knowledge of the value of $\cmin$.  Thus, before beginning
the {\it characterization} of the $\chi^2$ contour, we must {\it optimize}
by finding $\cmin$.

In the descriptions below, for each subsection we first give a short 
qualitative description to provide the key flavor of the subroutine, 
then a more quantitative description of its implementation, and finally 
any detailed treatments of computational subtleties. According to the 
readers' interest, they can go into whatever depth they want. The summary 
of the overall algorithm is given in Section~\ref{sec:final_algorithm}.

\subsection{Optimization in \ourCode} \label{sec:optimize}

The {\it optimization} carried out by \ourCode is based on the simplex
searching algorithm of \citet{simplex}.  It is well known that the simplex
search is not robust against functions with multiple disconnected local
minima.  We therefore augment the simplex with a Monte Carlo component designed
to allow \ourCode to dig itself out of false $\chi^2$ minima.

Qualitatively, the algorithm described below involves exploring the 
parameter space along diverse directions, detecting local minima, finding 
absolute minima, and checking for multimodality. In detail, for a parameter 
space of dimension $D$, 
initialize $2\times\left(D+1\right)+D/2$ points on an ellipsoid 
whose axis in each direction is the span of the full parameter
space to be searched.  As will be seen below, this initial search
culminates by using two independent Nelder-Mead simplexes to try
to find the global $\cmin$.  The number of points initialized is
chosen to be enough to initialize these simplexes (the
Nelder-Mead simplex is driven by the motion of $D+1$  `particles'
in parameter space) plus a little extra.
Each of these initial $2\times\left(D+1\right)+D/2$
particles will take $100\times D$ steps driven
by a Metropolis-Hastings MCMC algorithm with Gibbs sampling \citep{gibbs}
and simulated annealing \citep{annealing}.  

That is to say: at each step
for each particle, randomly choose a basis vector (i.e.\ direction in 
basis parameter space) along which to step.  Select
a value $r$ from a normal distribution with mean zero and standard deviation
unity.  Step from the particle's current position along the chosen basis vector
a distance $r\times n_i$ where $n_i$ is the difference between the maximum
and minimum particle coordinate value along the chosen basis vector.
Accept the step with probability
$\exp\left[-0.5\times(\chi^2_\text{new}-\chi^2_\text{old})/T\right]$ where 
the ``temperature'' $T$ is
initialized at unity and adjusted every $10\times D$ steps.  When adjusted, $T$ is
set to that value which would have caused half of the steps taken since the last
$T$-adjustment to be accepted.  Every $4\times D$ steps, select a new set of
random basis vectors and adjust the $n_i$ values accordingly.  Keep track of
both the absolute $\chi^2$ minimum and the recent $\chi^2$ minimum (to be defined
momentarily) encountered by each particle.  This is not necessarily
the particle's current position, as the $T$ term in the acceptance probability is
designed to cause the particles to jump out of local minima in search of other
local minima.  If ever a particle goes $10\times D$ steps without updating its
recent minimum, assume that it has settled into a local minimum from which
it will not escape. Move the particle back onto the ellipsoid where the particles
where initialized.  Set the particle's recent $\chi^2$ minimum (but not its absolute)
to the value at its new location (which will almost certainly not be a local minimum)
and resume walking.

After each particle has taken its $100\times D$ steps, choose the $D+1$ smallest
absolute minima encountered by the particles, and use them to seed a simplex
minimization search.  After that search, choose the $D+1$ smallest absolute
minima which are not connected to the absolute minimum with the lowest $\chi^2$
value and use them to seed another simplex search, in case the first simplex search
converged to a false minimum.  The connectedness of two absolute minima is judged
by evaluating $\chi^2$ at the point directly between the two minima.  If $\chi^2$
at this midpoint is less than the larger of the two $\chi^2$ absolute minima,
then it assumed that there is a valley in $\chi^2$ connecting the two, and they
are deemed ``connected.''  If not, it is assumed that there is a barrier of high
$\chi^2$ cutting the two off from each other, and they are deemed ``disconnected.''

This is a fairly weighty algorithm, requiring a number of $\chi^2$ evaluations that
scales as $D^2$.  However, in the case of highly complex likelihood surfaces, we find
that this is necessary for accuracy in the search for the absolute minimum 
as well as multimodality. In addition it increases efficiency in that it 
keeps \ourCode from wasting time characterizing the
$\chi^2$ contour around a false minimum.  We have tested this algorithm on several
multi-dimensional toy $\chi^2$ functions, running the algorithm many times on each function,
each time with a different random number seed.  We find that the current algorithm with 
its values tuned ($100\times D$ steps per particle, $4\times D$ steps before resetting
the bases, $10\times D$ steps before adjusting $T$, etc.) consistently finds the global
$\cmin$ or a point reasonably close to it. 
In fact we find in Sec.~\ref{sec:compare} 
that it frequently finds a lower $\cmin$ than MultiNest. 
That being said, we are open to the possibility that a
more efficient means of robustly optimizing $\chi^2$ exists and would welcome its
incorporation into \ourCode.  The algorithm described above is implemented in the
files \texttt{include/dalex\_initializer.h} and
\texttt{src/dalex/dalex\_initializer.cpp} in our codebase, for anyone wishing
to inspect it further.

\subsection{Characterization in \ourCode} \label{sec:characterize}

Once \ourCode has found $\cmin$, it can begin to characterize the shape of the
$\chi^2\le\clim$ contour surrounding that minimum.  This is achieved via
successive minimizations of the cost function introduced in equation~(\ref{eqn:cost}).
Equation~(\ref{eqn:cost}) is designed to drive
\ourCode to find points that are both inside the desired contour and far from points
previously discovered.  The neighbor function 
$N(\vec{\theta})$ increases as \ourCode gets farther
from previously known points. The exterior function $E(\chi^2)$ exponentially decays with $\chi^2$
once \ourCode leaves the confines of $\chi^2\le\clim$.  The parameter
$\ell$ in $E(\chi^2)$ can be tuned to allow \ourCode to take
circuitous routes through parameter space, swerving in and out of
$\chi^2\le\clim$ in search of the minimum of equation~(\ref{eqn:cost}).
We implement equation~(\ref{eqn:cost}) as a C\texttt{++} class \texttt{cost\_fn}
defined in the files \texttt{include/cost\_fn.h} and \texttt{src/dalex/cost\_fn.cpp} in
our codebase.  We describe that class below. 

\subsubsection{The cost function} \label{sec:cost}

In order to evaluate equation (\ref{eqn:cost}), one requires a set of known
$\chi^2\le\clim$ points from which to calculate the distances $N$.  Select a
set of previously discovered $\chi^2\le\cmin$ points $\{\vec{A}\}$ to 
represent the already explored region of the confidence
contour.  $\{\vec{A}\}$ will not always necessarily be all of the $\chi^2\le\clim$
points previously discovered by \ourCode.  If the full set of $\chi^2\le\clim$
points is large, it may be helpful to thin them, taking, for instance, every third
point and adding it to $\{\vec{A}\}$, to prevent
calculations of $N(\vec{\theta})$ from
taking too much time.  There may also be geometric considerations involved in what
goes into $\{\vec{A}\}$, which we will discuss below.  Upon instantiation,
\texttt{cost\_fn} loops through all of the points in $\{\vec{A}\}$, finding the
range of $\{\vec{A}\}$ in each of the cardinal basis directions.  Choose
one of these ranges, either the minimum or median as described below,
as the normalizing factor $s$.  When evaluating
equation~(\ref{eqn:cost}) at a point $\vec{\theta}$, 
the value of $N(\vec{\theta})$ is the harmonic mean of the parameter
space distances
\begin{equation}
\label{eq:nn_dist}
d_A=
\sqrt{\sum_i^D\left(\frac{\theta_i-\theta_{A,i}}{s}\right)^2}
\end{equation}
for all of the points $\vec{\theta}_A$ in $\{\vec{A}\}$. 

The purpose of the $N$ term in equation~(\ref{eqn:cost}) is to drive
simplex minimizations towards points in $\chi^2\le\clim$ that are far
from previously explored regions of parameter space.  The effectiveness
of our search, however, will still depend on where the simplex search
is initialized.  Do we start the simplex close to $\cmin$ and hope that
the repulsive properties of $N(\vec{\theta})$ will drive \ourCode
away from already explored points?  Or should we start the simplex
far outside of $\chi^2\le\clim$, and use the interplay between
the $\chi^2$ and $N(\vec{\theta})$ dependences of $F(\vec{\theta})$
to approach unexplored regions of $\chi^2\le\clim$ from without.
\ourCode uses both approaches complementarily, seeding the former
with the results of the latter as described below.

\subsubsection{Refining $\cmin$} \label{sec:refine}

First though, we add one refinement to the previous section. 
Since the utility of equation (\ref{eqn:cost}) involves 
correctly specifying $\clim$, \ourCode's performance depends on efficiently
finding the true $\cmin$ (this will be demonstrated in Section \ref{sec:limits} below).
Unfortunately, we sometimes find that in highly curved likelihood cases
even the elaborate
function minimization scheme described in Section \ref{sec:optimize} is insufficient
to find the true minimum $\chi^2$ value.  We therefore supplement Section~\ref{sec:optimize}
with the following extra algorithm to refine $\cmin$.

Initialize a set of $2\times D$ MCMC chains.  Take $4\times D$ steps
in each of these chains, driving the MCMC with $F(\vec{\theta})$ in place
of $\chi^2$.  Use simulated annealing \citep{annealing}
to keep the acceptance rate
near $0.5$.  After taking the $4\times D$ steps, randomly choose
$D$ of these particles.  Use the chosen particles and the current
$\cmin$ point to seed a simplex minimization of $\chi^2$.  If a new
$\cmin$ is found, repeat the process, remembering the positions of
the MCMC particles between iterations.
Once no new $\cmin$ is found, the refinement is considered complete. 
This is found to work very well; as mentioned previously, in difficult cases 
\ourCode routinely finds lower $\cmin$ than MultiNest does.


\subsubsection{Exploring $\chi^2\le\clim$ from without} \label{sec:explore}

With the $\cmin$ found, we are ready to characterize the confidence 
contour. The exploratory search described below is implemented by the
method \texttt{find\_tendril\_candidates()} defined in the
file \texttt{src/dalex/dalex.cpp} in our code base.

Our far-reaching simplex minimization of $F(\vec{\theta})$ works
as follows.  Fit a $D$-dimensional ellipsoid to all of the
$\chi^2\le\clim$ points discovered thus far, using the algorithm
described in Appendix~\ref{sec:ellipsoid}.  Initialize a cost function
using all of the $\chi^2\le\clim$ points discovered so far as
$\{\vec{A}\}$. If more than 20,000 $\chi^2\le\clim$ points have been
discovered so far, thin $\{\vec{A}\}$ so that it contains between
20,000 and 40,000 points; this is to make calculations of $N(\vec{\theta})$
tractable.  Set $\ell=0.25\times(\clim-\cmin)$ (minimum of $\ell=2.0$)
so that simplex minimization of $F(\vec{\theta})$
is free to use the region of parameter space just outside of $\chi^2\le\clim$
to avoid evaluating $\chi^2$ near already discovered regions of $\chi^2\le\clim$.
Set $s$ equal to the minimum range in the cardinal spans of $\{\vec{A}\}$ so
that $N(\vec{\theta})$ is dominant and will drive the simplex towards the far
edges of $\chi^2\le\clim$ as $F(\vec{\theta})$ is minimized.
For each of the $D$ axes of the ellipsoid above:

\begin{itemize}
\item (1E) Select a point $3\times r$ from the center of the ellipsoid,
where $r$ is the length of the semi-axis of the ellipsoid in the chosen direction.
\item (2E) Select $D$ other points centered on the point chosen in (1E).
These points will be $0.1\times r_i$ away from the point chosen
in (1E) along the other $D$ axes of the ellipsoid.
\item (3E) Use these $D+1$ points to seed a simplex
and minimize $F(\vec{\theta})$ from those points.  Add any $\chi^2\le\clim$
points discovered to $\{\vec{A}\}$.
\item (4E) Repeat steps (1E)-(3E) for the point $-3\times r$ away from the
ellipsoid's center.
\end{itemize}

In principle, this search begins by surrounding $\chi^2\le\clim$ at a great
distance (the initial seed points at $\pm 3\times r$ distance away from the
center of the ellipsoid) and closing in on the $\chi^2\le\clim$ contour,
hopefully finding the most distant corners of the contour under the influence
of $N(\vec{\theta})$.

Because the initialization of the cost function for this search requires
$\chi^2\le\clim$ points to populate $\{\vec{A}\}$, we initialize \ourCode
by running this search once with the initial seeds selected at $\pm r$
instead of $\pm 3\times r$.  This search is then immediately run again
with the usual seeds at $\pm 3\times r$.

\subsubsection{Exploring the cost function from within} \label{sec:tendril}

Though the algorithm described in Section \ref{sec:explore} is effective at
finding the extremities of $\chi^2\le\clim$, it has no mechanism for filling
in the complete contour.  Indeed, if $F(\vec{\theta})$ behaves the way it is
designed to behave, we should hope that the end result of Section
\ref{sec:explore} is small clusters of $\chi^2\le\clim$ points widely spread
from each other in parameter space with few evaluated points connecting them.
Therefore, in addition to the exploratory search of Section \ref{sec:explore},
\ourCode runs the following search to trace out the contour. 
We refer to this search as the ``tendril search,''
since it is designed to behave like a vine creeping across a wall. 
A tendril search is actually a series of simplex minimizations concatenated
in such a way as to force \ourCode to fully explore any non-Gaussian wings
present in the likelihood function.  A single \ourCode will perform several
independent tendril searches over the course of its lifetime.  We refer
to the individual simplex minimizations making up a single tendril search as
`legs' below. 

A tendril search is performed as follows.
After concluding the exploratory search described in Section \ref{sec:explore},
take the $2\times D$ end points of the simplex searches and rank them according
to their cost function values, using the cost function from Section
\ref{sec:explore}.  Save only the $D/2$ points with the lowest cost function values.
These will be the potential seeds for the tendril search.  When the time comes
to perform a tendril search, initialize a new cost function. Use every
$\chi^2\le\clim$ point so far discovered, except those discovered in the
course of the search described in Section \ref{sec:explore} to populate
$\{\vec{A}\}$.  Set $\ell=1.0$ and use the median cardinal span of $\{\vec{A}\}$
as the normalizing factor $s$. Rank the $D/2$ seed candidates chosen above
according to their values in the new cost function.  Choose the candidate
with the lowest cost function value, so long as that candidate is not inside
any of the exclusion ellipsoids $\{X\}$ (``exclusion ellipsoids'' will be defined
below; the first time the tendril search is run, $\{X\}$ is empty).  If no
candidates remain that are outside of $\{X\}$, re-run the search from
Section \ref{sec:explore}.  In this way, \ourCode alternates between trying to
fill in the $\chi^2\le\clim$ with a tendril search, and trying to expand the
boundaries of $\chi^2\le\clim$ with the exploratory search of
Section \ref{sec:explore}.

Every leg of a tendril search has an origin point $\vec{\gamma}$, from which
it will start, and a meta-origin point $\vec{o}$.  The meta-origin
$\vec{o}$ of the current leg is the origin $\vec{\gamma}$ of
the previous leg.  We use the vector $\vec{\gamma}-\vec{o}$
to impose structure on the seeds of the simplex underlying the leg
and thus try to keep the legs in a tendril search moving in the same
general direction and prevent them from
turning around before they have thoroughly explored their present region
of parameter space, assuming this still keeps them within
$\chi^2\le\clim$.  In the case of the first leg in a
tendril search, $\vec{o}$ is set to the $\cmin$ point.

Set the candidate selected from the end points of Section \ref{sec:explore}
as $\vec{\gamma}$ and build a simplex minimizer around it. 
The seeds for this simplex are found by
fitting an ellipsoid to all of the $\chi^2\le\clim$ points discovered by
every tendril search performed by \ourCode thus far
that are judged to be ``connected'' to $\vec{\gamma}$
(if this is the first tendril search, simply use all of the
$\chi^2\le\clim$ points discovered thus far).  Use bisection to find
$\chi^2=\clim$ points centered on $\vec{\gamma}$ but displaced from it
along the directions $\vec{e}+\vec{b}$ where $\vec{e}$ are the axial
directions of the ellipsoid fit to the $\chi^2\le\clim$ points and
$\vec{b}$ is the unit vector pointing parallel to $\vec{\gamma}-\vec{o}$.
Choose as seeds the $D$ points midway between $\vec{\gamma}$ and
the $\chi^2=\clim$ points discovered by bisection.  The $D+1$th seed point
is $\vec{\gamma} + \beta\vec{b}$ where $\beta$ is the average distance
from the first $D$ seed points to $\vec{\gamma}$.  Use these $D+1$ seed
points to seed a simplex minimization of the cost function initialized above.
Once the simplex minimization has converged, set
$\vec{\gamma}\rightarrow\vec{o}$ and set the endpoint of the simplex search to be
$\vec{\gamma}$ for the next leg of the tendril search. 
Rebuild the cost function adding any 
newly discovered $\chi^2\le\clim$ points to $\{\vec{A}\}$, and repeat the
process above.  In this way, \ourCode concatenates successive
simplex minimizations of $F(\vec{\theta})$, expanding the set of points used
to calculate $N(\vec{\theta})$ as it goes, hopefully filling in the
$\chi^2\le\clim$ contour as it proceeds.  This process is repeated until convergence,
which we will define below, at which point a new candidate is chosen from those
discovered in Section \ref{sec:explore}, and a new tendril search
is begun.

There are a few hitherto unmentioned subtleties in the tendril search as
described above.  The ellipsoid whose axial directions are used to find
the seeds for the simplex is fit to all of the $\chi^2\le\clim$
points deemed ``connected'' to $\vec{\gamma}$.  We defined one use of the
word ``connected'' in Section \ref{sec:optimize}: evaluate $\chi^2$ at the
point halfway between $\vec{\gamma}$ and $\vec{\theta}_i$.  If $\chi^2\le\clim$,
$\vec{\theta_i}$ is connected to $\vec{\gamma}$.  They are disconnected otherwise.
If this were really what \ourCode did to judge the connection between each point
evaluated in each tendril search, \ourCode would quickly blow its budget of $\chi^2$
evaluations on judging connectivity.  Therefore, as the tendril search progresses,
we keep a look up table associating each discovered $\chi^2\le\clim$ point with
a ``key point'' in parameter space.  For each individual leg of a tendril search, the
``key points'' are its origin, its end, and its midpoint.  Each $\chi^2\le\clim$
discovered during the search is associated with the key point that is closest
to it in parameter space.  To determine if a $\vec{\gamma}$ is connected to a
point previously discovered by tendril searching, it is deemed sufficient to
determine if $\vec{\gamma}$ is connected to the associated key point.  This
limits the number of unique $\chi^2$ calls used to judge connectivity.

While the tendril search as described thus far does a better job of
filling in the $\chi^2\le\clim$ contour than does the purely exploratory
search of Section \ref{sec:explore}, it is still focused on reaching out
into uncharted regions of parameter space.  As such, it can sometime miss
the breadth of the $\chi^2\le\clim$ contour perpendicular to the
$\vec{\gamma}-\vec{o}$ direction.
We therefore supplement the tendril search with the following search meant
to fill in the breadth of the contour regions discovered by pure tendril
searching.

After minimizing $F(\vec{\theta})$, find the vector $\vec{g}$ pointing from
$\vec{\gamma}$ to the endpoint of the simplex minimization.  Let $G$ be
the parameter space L2 norm of $\vec{g}$.  Select $D$ random normal vectors
$\vec{p}_i$ perpendicular to $\vec{g}$.  For each of the vectors
$\vec{p}_i$, construct a parameter space direction
\begin{equation}
\vec{c} = \vec{g}/G + \epsilon\vec{p}_i
\end{equation}
where $\epsilon$ is a random number between 0 and 1.  Normalize $\vec{c}$.
Sample $\chi^2$ at the points $\vec{\gamma} + \delta \vec{c}$, where $\delta$
steps from $0.1\times G$ to $G$ in $0.1\times G$ increments, thus constructing
a $D$-dimensional cone with its vertex at $\vec{\gamma}$ and opening towards
the end point of the simplex minimization.  In this way, \ourCode tries to
fill in the full volume of the region of $\chi^2\le\clim$ discovered by
the tendril search.  This search is performed after each leg of each
tendril search (i.e.\ after each individual simplex minimization
for a specific $[\vec{\gamma}, \vec{o}]$ pair).

We said above that \ourCode runs the tendril search ``until convergence''.
Convergence is a tricky concept in Bayesian samplers.  It is even trickier
in \ourCode. Because \ourCode is not attempting to sample a posterior
probability distribution, there is no summary statistic
analogous to Gelman and Rubin's $R$ metric for MCMC
\citep{GelmanRubin:1992, BrooksGelman:1998} that will tell us that
\ourCode has found everything it is going to
find.  We therefore adopt the heuristic convergence metric that as long as \ourCode
is exploring previously undiscovered parameter space volumes with $\chi^2\le\clim$,
\ourCode has not converged.  The onus is thus placed on us to determine in a way
that can be easily encoded which volumes in parameter space have already been
discovered.

In Appendix \ref{sec:ellipsoid} we describe an algorithm for taking
a set of parameter space points $\{\vec{P}\}$ and finding an approximately minimal
$D$-dimensional ellipsoid containing those points.  Each time \ourCode runs a
tendril search as described in the previous paragraphs, we
amass the $\chi^2\le\clim$ points discovered into a set $\{\vec{T}\}$ and fit an
ellipsoid to that set as described in Appendix \ref{sec:ellipsoid}.  While running
the tendril search, \ourCode keeps track of the end point of each individual leg of the
search.  If the simplex search ends in a point that is contained in one of the
ellipsoids resulting from a previous tendril search (the set of ``exclusion
ellipsoids'' $\{X\}$ referenced earlier), the simplex search is considered
to be doubling back on a previously discovered region of parameter space and a
``strike'' (analogous to what happens when a batter in baseball swings at the ball
and misses) is recorded.  Similarly, if the simplex
point ends inside of the ellipsoid constructed from the points $\{\vec{T}\}$ discovered
by the previous legs comprising the current tendril search (and the volume
of that ellipsoid has not expanded due to the current simplex) a ``strike'' is recorded.

Each time a ``strike'' is recorded, the tendril search backs up and starts from the
last acceptable simplex end point, adopting the corresponding
$\vec{\gamma}$ and $\vec{o}$.  If ever three ``strikes'' in a row are recorded, the
tendril is deemed to have converged (or, to continue the baseball analogy, to have
``struck out'').  An ellipsoid is fit to
the $\chi^2\le\clim$ points $\{\vec{T}\}$ discovered by the entire
tendril search.  This ellipsoid is added to the set $\{X\}$ of
previously discovered ``exclusion ellipsoids.''
By refusing to start a new tendril search in any
of the previously discovered exclusion ellipsoids, we increase the likelihood that
each successive sequence of tendril searches will discover a previously unknown region of
$\chi^2\le\clim$.

We find that in practice (see Section~\ref{sec:compare}) the ``three strikes
and you're out'' rule works well in both computational efficiency and
completeness in characterization -- often better than MCMC and MultiNest
algorithms that have declared convergence.

\subsubsection{Final algorithm} \label{sec:final_algorithm}

We have described three search modes in addition to the initial optimization of
Section \ref{sec:optimize}.  
Section \ref{sec:refine} describes an algorithm that takes steps which are
clustered around $\cmin$, trying to ensure that \ourCode has found the true value
of $\cmin$.  Section \ref{sec:explore} describes an algorithm
that searches for $\chi^2\le\clim$ regions by closing in on them from the
$\chi^2>\clim$ parameter space.  Section \ref{sec:tendril} explores the $\chi^2\le\clim$
contour from within, focusing on regions of parameter space that may not have been
explored yet. 
We find that the diversity of optimization and characterization steps 
is well suited, and indeed crucial, to dealing with complicated likelihood 
surfaces, with degenerate/highly curved/non-Gaussian or multimodal 
properties. Equally important, in high-dimensional cases, \ourCode 
converges in a number of $\chi^2$ evaluations significantly lower
than required by MultiNest 
while finding regions of parameter space ignored by MultiNest 
(and indeed often values of $\cmin$ lower than MultiNest). 

We combine these three search modes into the following master
algorithm for \ourCode.

\begin{itemize}
\item (1) Run the {\it optimization} algorithm from Section \ref{sec:optimize}.
\item (2) Perform the $\cmin$-refining search of Section \ref{sec:refine}.
\item (3) Perform a single sequence of tendril searches as described
in Section \ref{sec:tendril} (recall that if no viable candidates
for a tendril search starting point exist, we will perform the search in
Section \ref{sec:explore} to find some).
\item (4) Alternate steps (2) and (3) until convergence.
\end{itemize}

As stated before, ``convergence'' is not a well-defined formal concept 
for \ourCode. 
Presently, we have no global convergence condition.  \ourCode runs until it has
made a user-specified number of calls to $\chi^2$.  
Operationally, our local criterion works quite well, 
as shown in the next sections. Future work will explore 
development of a global convergence criterion.

\section{A Cartoon Likelihood Function} \label{sec:cartoon}

Algorithms such as MCMC, MultiNest, and \ourCode are designed for use in characterizing
likelihood functions complex enough that evaluating them the $\approx 100^D$ times
necessary to evaluate them directly on a well-sampled grid in parameter space 
would take too much
time to be feasible.  In many cases, such functions are so complex that even evaluating
them the (comparatively) few number of times necessary for the algorithms under
consideration to converge can take hours or, in the worst cases, days.
To facilitate more rapid testing during the development of \ourCode, we designed a cartoon
likelihood function which mimicked the non-linear, non-Gaussian behavior of realistic
multi-dimensional likelihood functions, but which could be evaluated almost instantaneously
on personal computers.  These functions are made available in our code via the header file
\texttt{include/exampleLikelihoods.h}.  The likelihood functions do depend on other
compiled objects from our code base.  Users should consult the
\texttt{Makefile}, specifically the example executables \texttt{curved\_4d},
\texttt{curved\_12d}, and \texttt{ellipse\_12d}, for
examples how to compile software using our toy likelihoods.

In order to understand how we have constructed our cartoon likelihood function,
it is useful to consider an example from nature.  Since the authors are cosmologists,
the example we choose is using the cosmic microwave background (CMB) 
anisotropy spectrum as measured by an experiment such as Planck 
\citep{planck16} to constrain the values of cosmological parameters.
The CMB likelihood function is, in the simplest case, defined on a 6-dimensional parameter space
$\{\Omega_m, \Omega_b, h, \tau, n_s, A_s\}$. 
(In reality there can be a dozen or more non-cosmological parameters as well.) 
Each combination of these 6 parameters implies a different spectrum of
anisotropy for the CMB.  This anisotropy is defined in
terms of the $C_\ell$ power coefficients of the multipole-moment $\ell$ 
decomposition of
the full CMB temperature distribution on the sky. Typical CMB experiments 
measure a few thousand of these $C_\ell$ coefficients (in the temperature-temperature
correlation function).  To compute a likelihood value for any combination of
cosmological parameters, the parameters are converted into their predicted
$C_\ell$ spectrum and this $C_\ell$ spectrum is compared to the actual
measurement.  In other words, there is an unknown, non-linear
function $g(\vec{\theta}, \ell)$ such that
\begin{equation}
\label{eqn:analogy}
g\left(\{\Omega_m, \Omega_b,h, \tau,n_s,A_s\}, \ell\right)=C_\ell
\end{equation}
To find the likelihood of any given combination of cosmological parameters,
one must first evaluate $g(\vec{\theta}, \ell)$ to get a test set of $C_\ell$'s
and then compare this set to the actual measured $C_\ell$'s, taking into account
the appropriate covariance matrix.

To mimic this behavior, we construct our cartoon likelihood functions in
three steps.

\begin{itemize}
\item (1) A function $\tilde{f}(\vec{\theta})=\vec{\mu}$ maps the cardinal
parameters $\vec{\theta}$ into auxiliary parameters $\vec{\mu}$.

\item (2) A second function $\tilde{g}(\vec{\mu}, \vec{x}) = \vec{y}(\vec{x})$
maps the parameters $\vec{\mu}$ and an independent variable $\vec{x}$ (analogous
to $\ell$ in the cosmological case (\ref{eqn:analogy}) above) to a dependent
value $\vec{y}$ (analogous to $C_\ell$). 

\item (3) A noisy cartoon data set $(\vec{x}, \vec{y}_0, \sigma_{\vec{y},0})$
is constructed by evaluating
\begin{equation}
\vec{y}_\text{mean}=\tilde{g}\left(\tilde{f}\left(\vec{\theta}_0\right), \vec{x}\right)
\end{equation}
at 100 values of $x$.  At each of these values of $(x, \vec{y}_\text{mean})$,
100 samples are drawn from a Gaussian distribution centered on $\vec{y}_\text{mean}$.
(A Gaussian distribution is chosen as a simplification of experiment noise 
properties for the purposes of rapid testing.) 
$\vec{y}_0$ is the mean of these 100 samples and $\sigma_{\vec{y},0}$ is the
standard deviation of these samples.
\end{itemize}
To evaluate $\chi^2$ at a test value of $\vec{\theta}$, one simply evaluates
\begin{equation}
\vec{y}_\text{test}=
\tilde{g}\left(\tilde{f}\left(\vec{\theta}_\text{test}\right), x\right)
\end{equation}
and compares
\begin{equation}
\label{eq:chi2_illustration}
\chi^2\left(\vec{\theta}_\text{test}\right)
=\sum_i^{100}\left[\left(\vec{y}_{\text{test},i}-\vec{y}_{0,i}\right)/\sigma_{\vec{y},0,i}\right]^2
\end{equation}

\subsection{Confidence versus Credible Limits: How to Set $\clim$}
\label{sec:limits}

Before proceeding, we have to define exactly what contours we are searching for. 
(The reader more interested in a direct comparison of results from 
different algorithms can skip to Section~\ref{sec:compare}.) 
A significant philosophical difference between the sampling algorithms
(MCMC and MultiNest) and the search algorithms (APS and \ourCode) is in how
each defines the desired boundary in parameter space (the ``confidence limit'' in
frequentist terminology; the ``credible limit'' for Bayesians).
Sampling algorithms learn the definition
boundary as they go.  They draw samples from the
posterior probability distribution and, after convergence has been achieved,
declare the limit to be the region of parameter space containing the desired
fraction of that total distribution.  Search algorithms require the user to
specify some $\clim$ defining the desired limit a priori.  This can be done
either by setting an absolute value of $\clim$ or by demanding that $\chi^2$
be within some $\Delta\chi^2$ of the discovered $\cmin$, i.e.
$\clim=\cmin+\Delta\chi^2$.  These two approaches are not wholly irreconcilable.
Wilks' Theorem \citep{wilks} states that, for approximately Gaussian likelihood
functions, the Bayesian credible limit (i.e. the
contour bounding $(1-\alpha)\%$ of the total posterior probability) in a
$D$-dimensional parameter space is equivalent to the contour containing all
points corresponding to $\chi^2\le\chi^2_\text{min}+\chi^2_{(1-\alpha)\%}$
where $\chi^2_{(1-\alpha)\%}$ is the value of $\chi^2$ bounding the desired
probability for a $\chi^2$ probability distribution with $D$ degrees of freedom.
Even given this result, there remains some ambiguity in how one sets
$\chi^2_{(1-\alpha)\%}$.  Namely: one must choose the correct value of $D$
for the degrees of freedom.  Traditionally, credible and confidence limits are
plotted in 2-dimensional sub-spaces of the full parameter space.  The limits
that are plotted are either marginalized or projected down to two dimensions.
In the case of marginalization, the Wilks'-theorem equivalence between confidence
and credible limits holds for $D=2$ degrees of freedom and $\Delta\chi^2=6.0$ 
for the 95\% contour. 
In the case of projections, $D$ is equal to the full dimensionality of the parameter
space. We argue and demonstrate below that the preferred choice,
with the most robust results, is projection with
$D$ equal to the full dimensionality of the parameter space being considered.
We illustrate this with a toy example.

\begin{figure}
\includegraphics[scale=0.2]{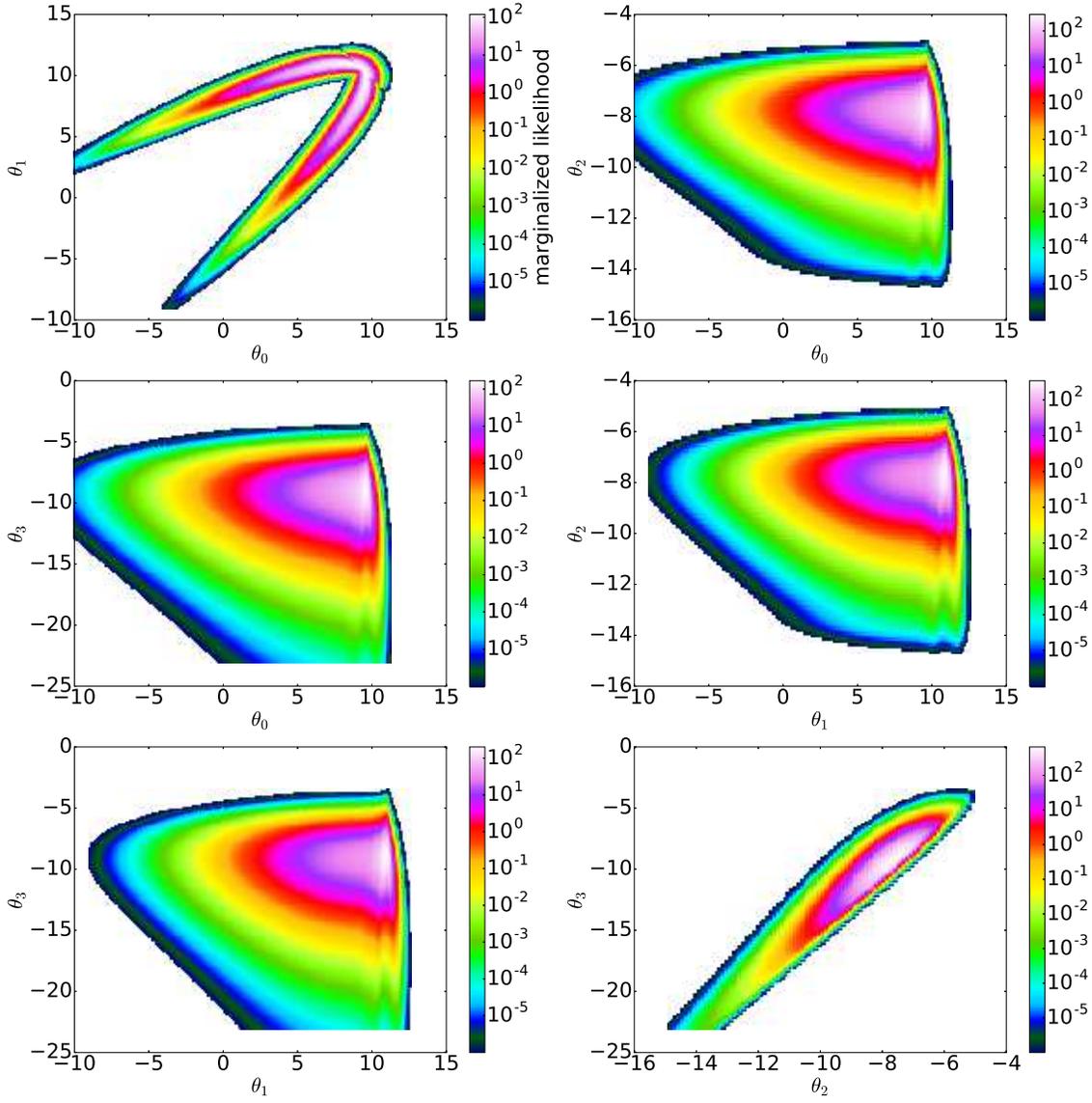}
\caption{
The marginalized likelihood in all of the 2-D sub-spaces
of our 4-D cartoon likelihood function is shown, with the color bar indicating 
the marginalized likelihood.  Note: the apparent discontinuity in the
shape of the confidence contour near the bend in the $\{\theta_0,\theta_1\}$
sub-space is an actual feature of our likelihood function, resulting from 
how the function is implemented in C\texttt{++}. 
}
\label{fig:cartoon_heat_map}
\end{figure}

In order to compare different definitions of confidence and credible limits, we
construct a cartoon likelihood function on 4 parameters.  Choosing such a
low-dimensional parameter space allows us to grid the full space and directly
integrate the Bayesian credible limits without requiring unreasonable
computational resources.  Figure \ref{fig:cartoon_heat_map} shows the marginalized
likelihood of our 4-dimensional cartoon likelihood function in each of the 2-dimensional
parameter sub-spaces.  As one can see, we have constructed our cartoon likelihood function
to have a severely curved degeneracy in the $\{\theta_0, \theta_1\}$ parameter sub-space and
significantly non-Gaussian contours in the other five sub-spaces.

\begin{figure}
\includegraphics[scale=0.9]{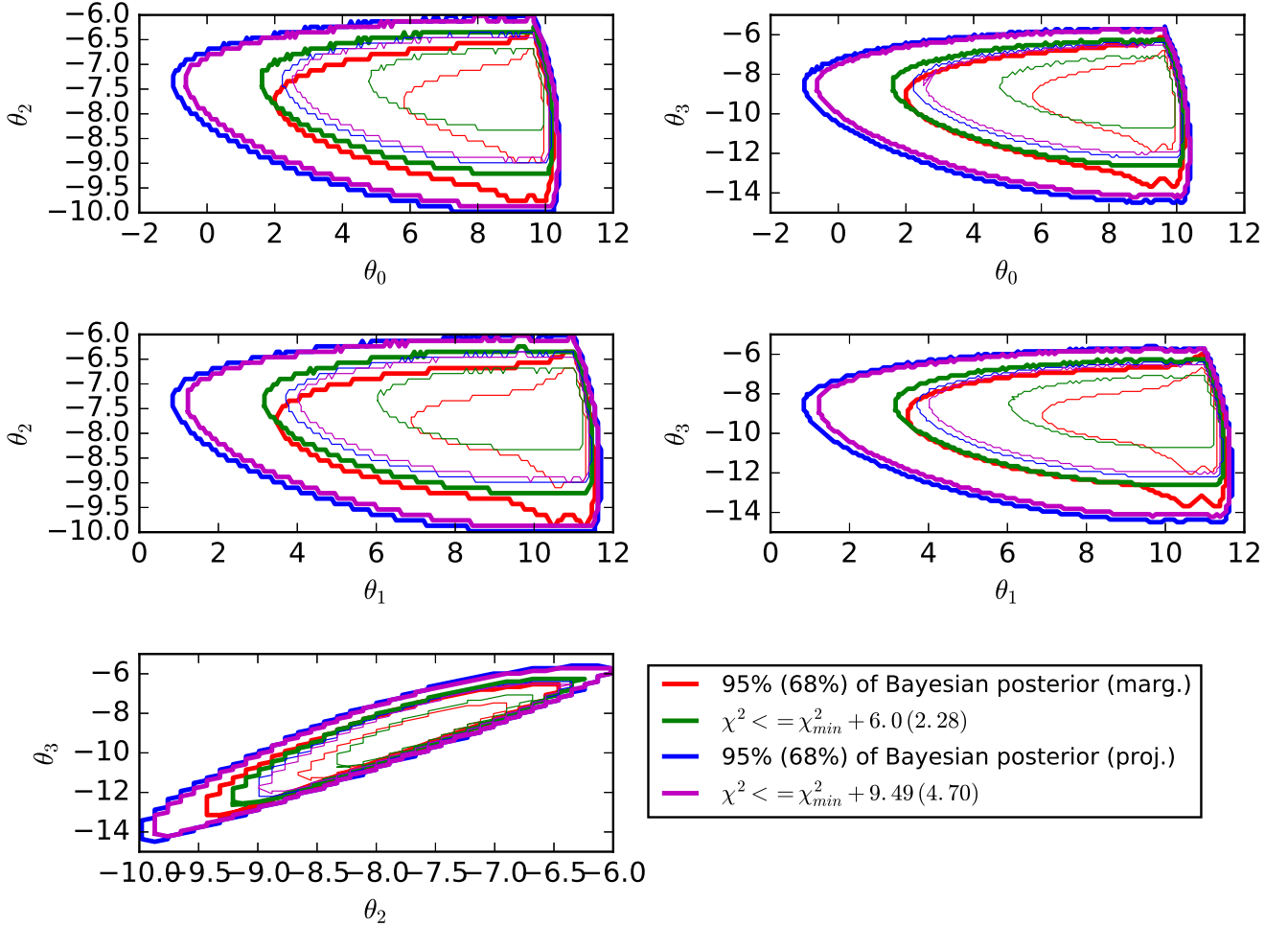}
\caption{
Comparison of the different confidence and credible limits are shown for the likelihood function
introduced in Figure \ref{fig:cartoon_heat_map}. 
The {\it projected\/} Bayesian credible limit 
agrees much better with the frequentist full dimensional likelihood ratio test 
($\chi^2$) confidence limit than the {\it marginalized\/} Bayesian 
credible limit. Thick contours are the 95\% limit; 
thin contours are the 68\% limit.  
The $\{\theta_0, \theta_1\}$ parameter sub-space
is shown in Figure \ref{fig:cartoon_truth_0_1}. 
}
\label{fig:cartoon_truth}
\end{figure}

\begin{figure}
\includegraphics[scale=0.9]{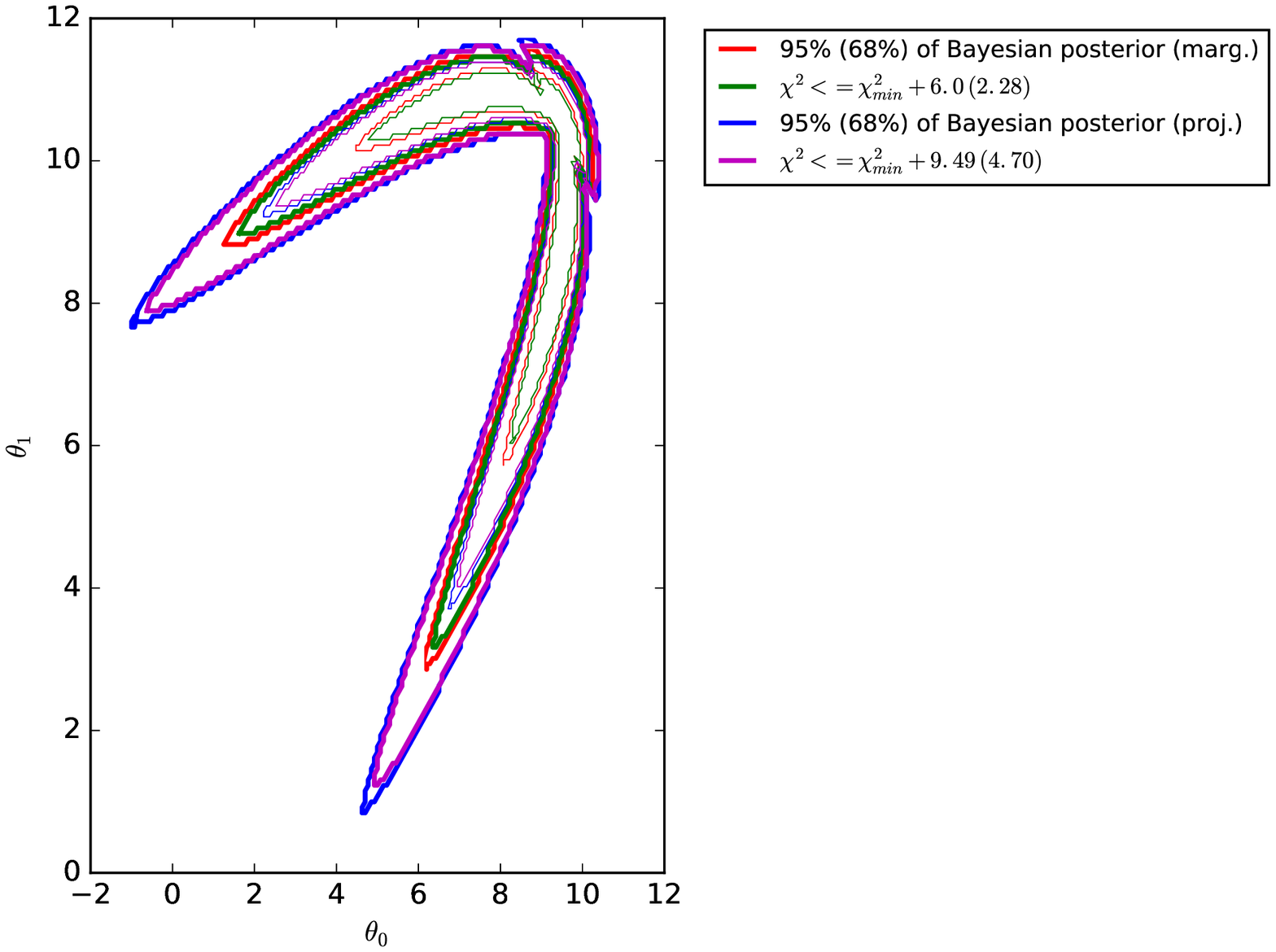}
\caption{
As Figure~\ref{fig:cartoon_truth} but for the $\theta_0$-$\theta_1$ subspace. 
Due to the better Gaussianity in this subspace, both the projected and 
marginalized Bayesian credible limits agree with the likelihood ratio test. 
}
\label{fig:cartoon_truth_0_1}
\end{figure}

In Figures~\ref{fig:cartoon_truth} and \ref{fig:cartoon_truth_0_1}, we show
the confidence and credible limits of our 4-dimensional toy likelihood function
computed in several ways.  The red contours show the {\it marginalized\/} Bayesian
credible limits.  These are the contours enclosing 95\% (68\%) of the marginalized
likelihood shown in Figure \ref{fig:cartoon_heat_map}.  The green contours are
the projections of the 4-dimensional contours enclosing
$\chi^2\le\chi^2_\text{min}+6.0\ (2.28)$, i.e. they were generated by finding all of the
points in 4-dimensional space satisfying $\chi^2\le\chi^2_\text{min}+6.0\ (2.28)$,
converting those points into a scatter-plot in each of the 2-dimensional sub-spaces, and
then drawing a contour around those scatter plots.  Naively, one would expect these
contours to be equivalent to the marginalized Bayesian contours, since 6.0 (2.28)
is the 95\% (68\%) limit for a $\chi^2$ distribution with two degrees of freedom.
(Note that the 95.4\% and 68.3\% limits, i.e.\ $2\sigma$ and $1\sigma$ in Gaussian 
probability, would have $\Delta\chi^2=6.17$ and 2.30.) 
This, however, is clearly not the case.  The non-Gaussian nature of our toy likelihood
prevents Wilks' theorem from applying in the marginalized case.  While you can integrate
away $d$ dimensions of a $D$-dimensional multivariate Gaussian and come up with a
$D-d$ multivariate Gaussian, no such guarantee exists for arbitrary, non-Gaussian functions.

The blue contours in Figure \ref{fig:cartoon_truth} and \ref{fig:cartoon_truth_0_1} show
the {\it projected\/} Bayesian credible limits of our cartoon function.
These were constructed by assembling all of the points in 4-dimensional space that contained
95\% (68\%) of the total likelihood, converting those points to a scatter plot in the
2-dimensional sub-spaces, and drawing a contour around the scatter plots.  The purple
contours show the similarly projected contours containing all of the points for which
$\chi^2\le\chi^2_\text{min}+9.49\ (4.70)$, with 
9.49 (4.70) being the 95\% (68\%) limit for
a $\chi^2$ distribution with four degrees of freedom.  We refer to these contours as
the ``full-dimensional LRT (likelihood ratio test)'' contours.
We see remarkably good agreement between the projected Bayesian and full-dimensional
LRT contours in both the 95\% and the 68\% limit, in accordance with
Wilks' theorem (since we have not marginalized away any of the 4 parameters in our
parameter space, Wilks' theorem still appears to hold). 

In the rest of this paper, we will examine likelihood functions based on their
projected Bayesian and full-dimensional LRT contours.  Though it is traditional
to consider marginalized contours in 2-dimensional sub-spaces, we note concerns that this
may discard important information about the actual extent of parameter space
considered reasonable by the likelihood function.  In all cases in Figures
\ref{fig:cartoon_truth} and \ref{fig:cartoon_truth_0_1}, the projected Bayesian
and full-dimensional LRT contours were larger than their 2-dimensional marginalized
counterparts.  This should not be surprising.  The projected Bayesian and full-dimensional
LRT contours contain all of the pixels in the full 4-dimensional space that fall within
the desired credible limit.  The marginalized counterparts take those full 4-dimensional
contours and then weight them by the amount of marginalized parameter space volume
they contain.  The final marginalized contour thus represents an amalgam of the raw
likelihood function and its extent in parameter space.  It is not obvious that this is
necessarily what one wants when inferring parameter constraints from a data set.  

If a point in
parameter space falls within the projected Bayesian or full-dimensional LRT contour,
it means that the fit to the data provided by that combination of parameters is,
by some metric, reasonable. This is somewhat a matter of taste, but 
if one is really interested in all of the parameter
combinations that give reasonable fits to the data, the conservative approach would be to keep
all such combinations, rather than throwing out some, simply because they include rare -- but fully accepted by the data -- 
combinations of, say, $\{\theta_1, \theta_3\}$ (in the case of the gap between the
marginalized and projected Bayesian contours in the $\{\theta_0, \theta_2\}$ plot in
Figure~\ref{fig:cartoon_truth}). 
For this reason, we proceed\footnote{ 
We do not consider useful the strictly frequentist
perspective that, since our cartoon likelihood function ``measures'' $\vec{y}(x)$
at 100 values of $x$, 
the confidence limit 
limit should therefore be set according to the $\chi^2$ distribution with 
100 degrees of freedom. 
This would set the 95\% confidence limit at $\chi^2=124.35$. 
Note that since $\chi^2_\text{min}$ for our toy likelihood
function ended up being 83.12, such frequentist confidence limit contours 
would be significantly
larger than anything shown in Figures~\ref{fig:cartoon_truth} and \ref{fig:cartoon_truth_0_1}.
Such large contours seem unlikely to be useful.
}, 
comparing the performance
of \ourCode with MultiNest by examining the projected Bayesian and full-dimensional LRT
contours.

We acknowledge that there are some applications for which the approximate equivalence of
$\chi^2\le\clim$ with a Bayesian credible limit for which we have argued above is insufficient. 
If a likelihood function is sufficiently non-Gaussian or a data set is sufficiently noisy, the
two will not be equivalent.  Indeed, in Figures \ref{fig:cartoon_truth} and
\ref{fig:cartoon_truth_0_1}, the blue and purple contours are not identical, but merely
approximately the same.  In cases where only the Bayesian credible limit is acceptable, we
still believe that there is a use for \ourCode.  Bayesian sampling algorithms generally 
get bogged down during ``burn-in'', the period of time at the beginning of the sampling run during
which the algorithm is simply learning where the minimum $\chi^2$ point is and what the
degeneracy directions of the contour are without efficiently sampling from the posterior.  We
show below that \ourCode converges to its confidence limits, with the same $\cmin$ and degeneracy
directions as the Bayesian credible limit, irrespective of one's opinion of Bayesian versus
frequentist statistical interpretations, after an order of magnitude fewer calls to $\chi^2$
than required by MultiNest.  If one does not feel inclined to trust the full confidence limits
returned by \ourCode, one should be able to achieve a significant speed-up in Bayesian
sampling algorithms by using \ourCode as pre-burner: defining the region of parameter
space to be sampled before sampling commences.

\section{Comparison of Techniques}
\label{sec:compare}

As a proof of concept, we begin by comparing the performance of
MultiNest and \ourCode on the 4-dimensional likelihood function
presented in Figure \ref{fig:cartoon_heat_map}.  As said before,
four is an extremely low dimensionality and we find that MultiNest
converges thoroughly and much more rapidly than \ourCode (though 
MultiNest does include some spurious points compared to exact, 
brute force calculation, unlike \ourCode, i.e.\ 
it has lower purity).  Later, we
will see that \ourCode out-performs MultiNest on higher-dimension
likelihood functions. The purpose of this current discussion is
to demonstrate that \ourCode does, in fact, find the limits expected
from Figures \ref{fig:cartoon_truth} and \ref{fig:cartoon_truth_0_1}. 

Comparing MultiNest and \ourCode requires defining how long each should 
run. 
Because MultiNest is a sampling algorithm designed to integrate
the posterior distribution, it has a well defined convergence criterion:
when the Bayesian evidence integrated over parameter space ceases to
grow, the sampling is done.  \ourCode does not have any analogous global
convergence criterion.  Individual tendril searches within \ourCode can
be said to converge when they ``strike out'' (see Section \ref{sec:tendril})
but this says nothing about the overall convergence of \ourCode.  It is up to
the user to specify how many $\chi^2$ evaluations they wish \ourCode
to make.  In Figure~\ref{fig:compare_4d_0_1}
below, we wish to show the convergence of MultiNest and \ourCode as
a function of ``time'' (time being measured in the number of $\chi^2$
evaluations being made).  To do this, we initialized several MultiNest
runs, each with a different number of ``live'' points.  MultiNest instances
with more live points took longer to converge 
(but can be expected to evaluate the posterior distribution more 
accurately). 
We then ran a single instance
of \ourCode and asked it to quit after 40,000 $\chi^2$ evaluations.
In Figure \ref{fig:compare_4d_0_1}, we
compare the MultiNest runs with the progress made by \ourCode after it
had evaluated $\chi^2$ as many times as it took MultiNest to converge. 
The red points are the 2-dimensional projections of the points discovered
by MultiNest to be in the 95\% Bayesian credible limit.  The blue points
are the $\chi^2\le\chi^2_\text{min}+9.49$ points discovered by \ourCode
at the equivalent point in time.  As stated, 
\ourCode does indeed find the correct $\chi^2\le\clim$
contour while MultiNest is faster for $D=4$ and somewhat impure. 
We show below that, in higher-dimensional cases, \ourCode characterizes 
the contour much more robustly than MultiNest.  We used version 3.9 of
MultiNest here and throughout this paper.  The most up-to-date version
of MultiNest can be downloaded from
\url{https://ccpforge.cse.rl.ac.uk/gf/project/multinest/}

\begin{figure}
\includegraphics[scale=0.9]{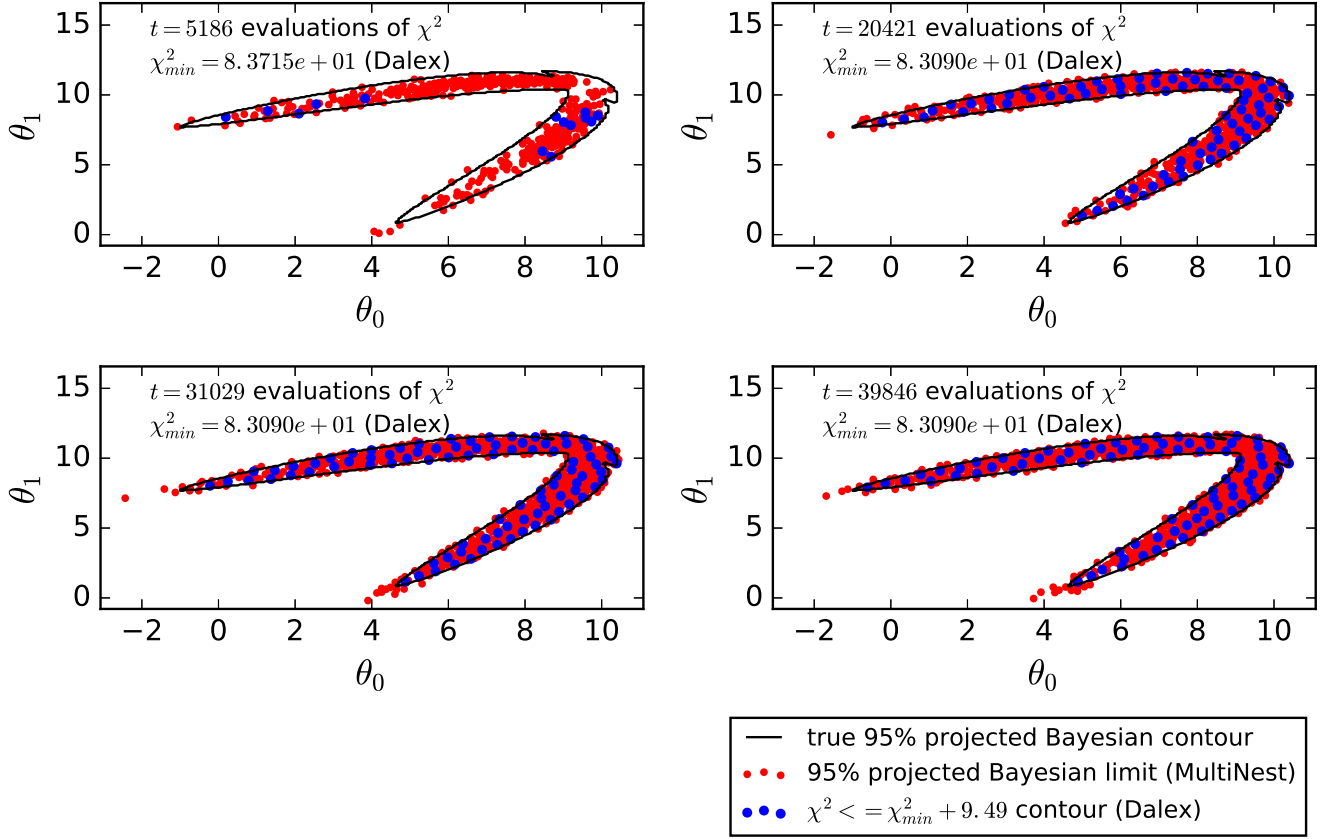}
\caption{
We demonstrate the convergence properties of MultiNest
and \ourCode on the 4-dimensional likelihood function from
Figure \ref{fig:cartoon_heat_map}.  The black contour represents
the true projected 95\% Bayesian credible limit.  The red points
represent the points in the projected 95\% Bayesian credible
limit discovered by MultiNest.  The blue points represent the
$\chi^2\le\chi^2_\text{min}+9.49$ points discovered by \ourCode.
Each panel represents a different MultiNest run, which took $t$
points to converge.  The blue points are all drawn from a single
run of \ourCode at the point during its history at which it had
evaluated $\chi^2$ $t$ times.
}
\label{fig:compare_4d_0_1}
\end{figure}

\subsection{Comparison on 12-dimensional case} \label{sec:12d}

We showed above that \ourCode found similar confidence limits,
but took much longer to converge than did MultiNest on a
4-dimensional likelihood function.  Now we compare \ourCode to MultiNest
on a 12-dimensional cartoon likelihood function.  Here we find that \ourCode
converges an order of magnitude faster than MultiNest.

\begin{figure}
\includegraphics[scale=0.9]{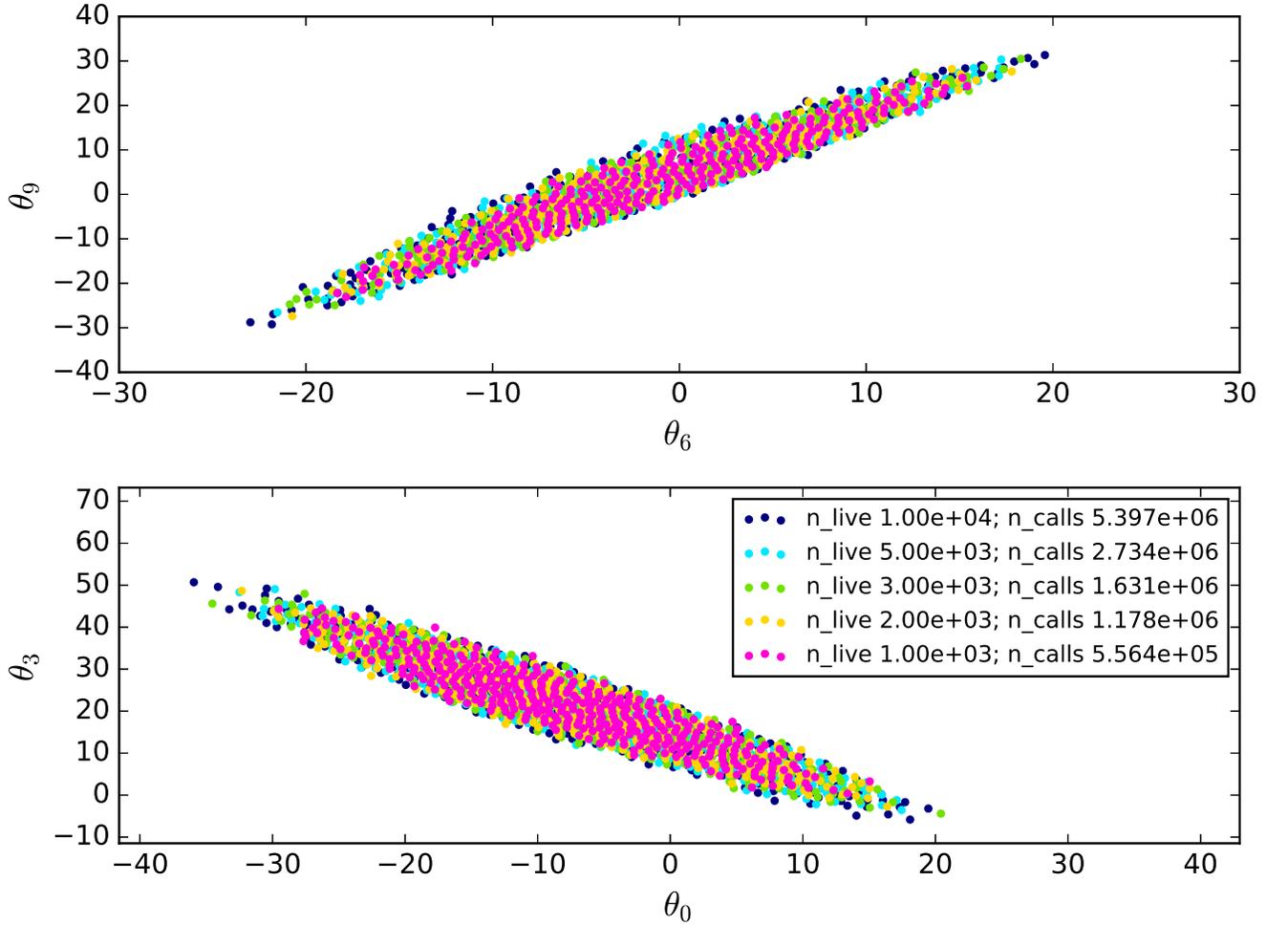}
\caption{
The projected 95\% Bayesian credible limit found by MultiNest on
one of our 12-dimensional cartoon likelihood functions.  Each color denotes a different
instance of MultiNest run with a different number of `live' points (denoted
by \texttt{n\_live} in the legend).  \texttt{n\_calls} indicates how many
calls to the likelihood function were necessary before MultiNest declared
convergence.
}
\label{fig:lump_nlive}
\end{figure}

\begin{figure}
\includegraphics[scale=0.9]{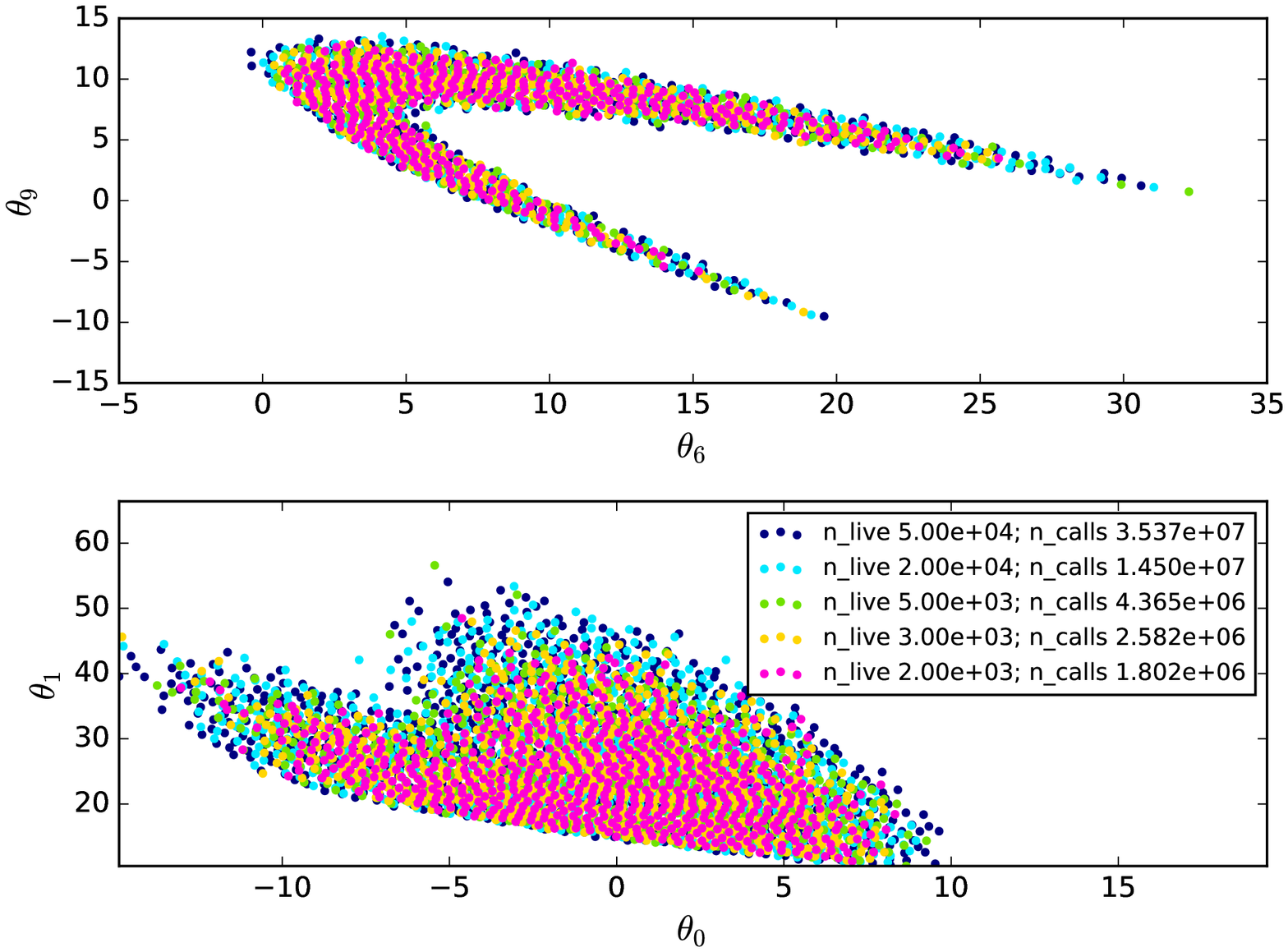}
\caption{
As Fig.~\ref{fig:lump_nlive} but for the curved degeneracy likelihood 
function. 
}
\label{fig:jellybean_nlive}
\end{figure}

Before we show the direct comparison between \ourCode and MultiNest, we must
be careful to define what we are comparing.  MultiNest involves several
user-determined parameters that control how the sampling is done.  One
in particular, the number of `live' points, which is somewhat analogous
to the number of independent chains run in an MCMC, has a profound effect
both on how fast MultiNest converges and what limit it finds when it does
converge.  We created two 12-dimensional instantiations of our cartoon
likelihood function, one with a very curved parameter space degeneracy,
one without, and ran several instantiations of MultiNest with different
numbers of `live' points on those functions.  Figures~\ref{fig:lump_nlive}
and \ref{fig:jellybean_nlive} show the projected Bayesian 95\% credible
limits found by MultiNest in two 2-dimensional sub-spaces for each
of those likelihood functions as a function of the number of `live' points.
The figures also note how many evaluations of the likelihood function were
necessary before MultiNest declared convergence.  Not only does increasing
the number of `live' points increase the number of likelihood evaluations
necessary for MultiNest to converge, it also expands the discovered credible
limits.  We interpret this to mean that running MultiNest with a small number
of `live' points results in convergence to an incomplete credible limit.
When comparing \ourCode to MultiNest, we must therefore be careful which
instantiation of MultiNest we compare against.  Selecting a MultiNest run with
too few `live' points could mean that MultiNest appears as fast as \ourCode, 
but converges to an incorrect credible limit.  Selecting a MultiNest run with
too many `live' points ensures that MultiNest has converged to the correct
credible limit, but in an artificially large number of likelihood evaluations.
In Figures \ref{fig:lump_0_3} and \ref{fig:lump_6_9}, 
and Figures~\ref{fig:jellybean_0_1} 
and \ref{fig:jellybean_6_9} below, we compare
\ourCode both to a MultiNest instantiation that we believe has converged to
the correct credible limit and a MultiNest instantiation that converges in a
time comparable to \ourCode.  This is to illustrate the relationship between
convergence time and accuracy in MultiNest.

In Figures \ref{fig:lump_0_3} and \ref{fig:lump_6_9} we compare the convergence of
\ourCode with that of MultiNest as a function of time.  The likelihood function
considered is a 12-dimensional cartoon constructed as in
Section \ref{sec:cartoon}, but without the highly curved degeneracy direction.
Each panel in the figures shows the state of the 95\% confidence
limits discovered by \ourCode after a specified number of evaluations of the
likelihood function.  The limits are plotted as scatter plots of the $\chi^2\le\clim$
points discovered by \ourCode at that point in time.  Because MultiNest does not produce
chains in the same way that
MCMC does, we cannot display the evolution of the credible limit found by MultiNest
as a function of time.  Instead, we plot two credible limits: one found by an
instantiation of MultiNest that converges after $210,000$  evaluations of
the likelihood function and one that converges after $2,700,000$ evaluations
of the likelihood function.  
As explained above, this difference in convergence time
was achieved by tuning the number of `live' points in the MultiNest run.  There are
two important features to note here.

The first feature is that \ourCode has converged to its final credible limit 
after about 125,000-150,000 calls to $\chi^2$.  The second feature is the difference
between the credible limits discovered by the two instantiations of MultiNest.
True, one of the MultiNest instantiation converges in roughly (within a factor of
two) as many likelihood evaluations as \ourCode.  This instantiation of MultiNest,
however, fails to explore the full non-Gaussian wings of the credible limit. 
See Figure~\ref{fig:lump_1d} for a visualization of the non-Gaussianity. 
The slower MultiNest instantiation does find the same limits as \ourCode.  However,
it requires more than an order of magnitude more likelihood function evaluations to
do so.  This is the origin of our claim that \ourCode converges an order of magnitude
faster than MultiNest on high-dimensional likelihood functions.  \ourCode does
not do as well as MultiNest characterizing the full width of the contour in
the narrow degeneracy direction in Figure \ref{fig:lump_6_9}. However, 
by the end of the full 250,000 $\chi^2$-evaluation
run, one can see the wings of the contour beginning to trace out the full boundary.
(Alternately, if one adopted the strategy suggested at the end of Section~\ref{sec:limits},
using \ourCode to set the proposal distribution for a sampling algorithm, one
could imagine filling in these contours more thoroughly than with just MultiNest alone
while still only requiring $\sim$ 10\% more evaluations of the likelihood function.) 

Figures \ref{fig:jellybean_0_1} and \ref{fig:jellybean_6_9} show a similar
comparison in the case of a 12-dimensional likelihood function with a highly curved
parameter space degeneracy.  Here again we see that \ourCode finds essentially the
same limits as MultiNest, but more than an order of magnitude faster.  We have
done preliminary tests with a 16-dimensional likelihood function.  These show 
almost two orders of magnitude speed-up of \ourCode vis-\'a-vis MultiNest.
In the 16-dimensional case, MultiNest requires 75 million calls to $\chi^2$
to converge. \ourCode finds the same credible limit after only 1 million calls
to $\chi^2$.

Regarding some computational specifics, note we 
have been measuring the speed of \ourCode and MultiNest in terms of how many
calls to $\chi^2$ are required to reach convergence.  This is based on the
assumption that, for any real physical use of these algorithms, the evaluation
of $\chi^2(\vec{\theta})$ will be the slowest part of the computation.  For
MultiNest, this is a fair assumption.  As a Markovian process, the present
behavior of MultiNest does not depend on its history.  The same statement is not
true of \ourCode.  Because \ourCode is attempting to evaluate $\chi^2$ at
parameter-space points as far as possible away from previous $\chi^2$
evaluations, \ourCode must always be aware of the full history of its search. 
Thus, the \ourCode algorithm implies some computational overhead in addition to
the expense of simple evaluating $\chi^2$.  For our 12-dimensional test cases,
this overhead averages out to between 0.005 and 0.015 seconds per $\chi^2$
evaluation on a personal laptop with a 2.9 GHz processor.  For the
16-dimensional test case referenced above, this overhead varied between
0.01 and 0.03 seconds per $\chi^2$ evaluation on the same machine.  It is
possible that future work with an eye towards computational optimization could
reduce these figures.  It is also possible that the steady march of hardware
innovation could render such concerns irrelevant going forward.  The memory
footprint of \ourCode appears to be well within the capabilities of modern
personal computers.

\begin{figure}
\includegraphics[scale=0.9]{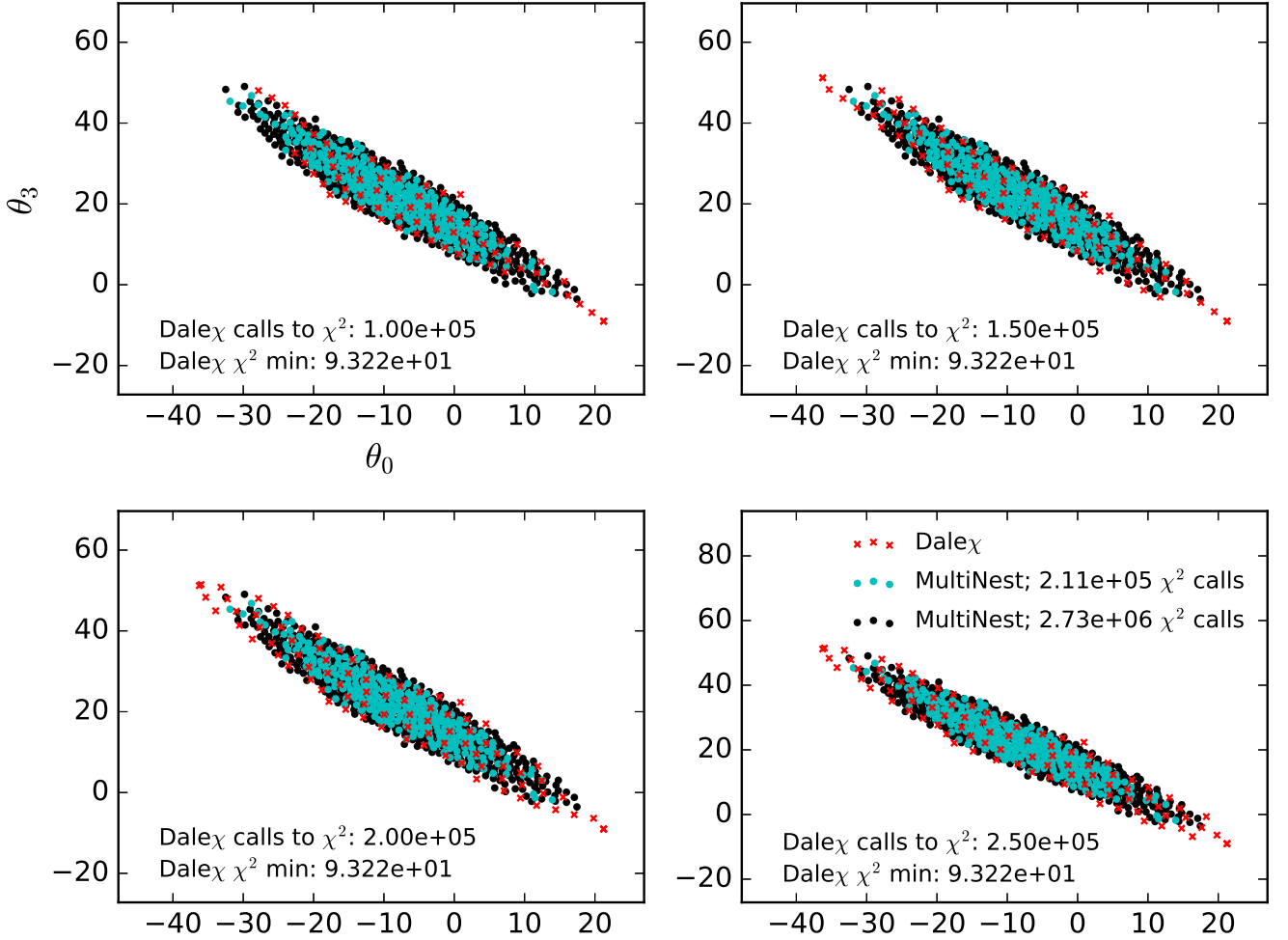}
\caption{
We compare the performance of \ourCode to MultiNest on a cartoon
likelihood function constructed as described in Section \ref{sec:cartoon}.
This cartoon is a 12-dimensional likelihood function in which several of the
dimensions have non-Gaussian posteriors (see the marginalized one-dimensional
posteriors presented in Figure \ref{fig:lump_1d}).  Each frame shows the progress of \ourCode 
after a specified number of calls to $\chi^2$.  The black and cyan points
indicate the projected Bayesian 95\% credible limit discovered by MultiNest
after a specified number of likelihood evaluations and are the same in 
all frames.  Note the difference
between the limits discovered by the rapid convergence MultiNest and the
slow convergence MultiNest.
}
\label{fig:lump_0_3}
\end{figure}

\begin{figure}
\includegraphics[scale=0.9]{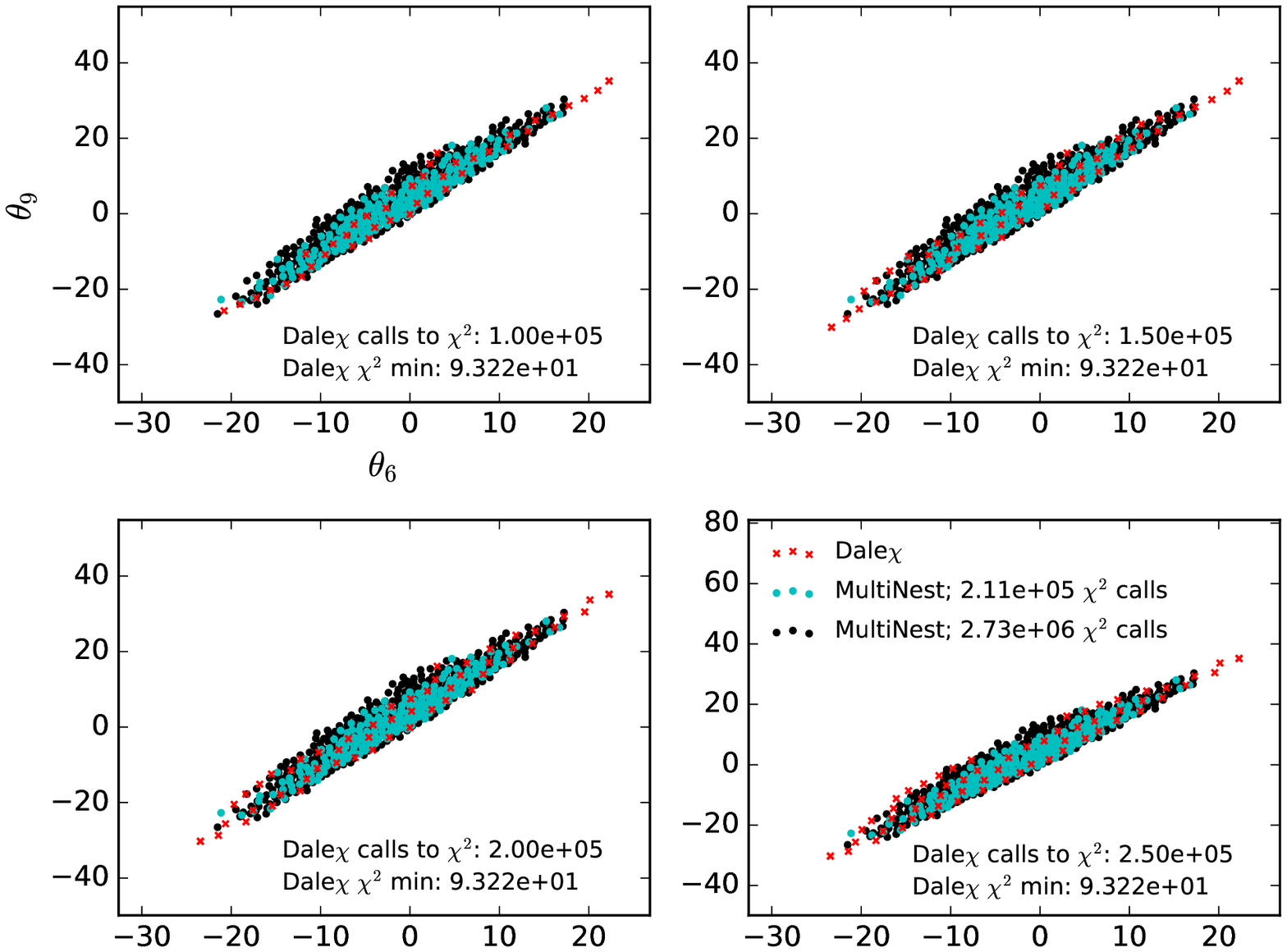}
\caption{
As Fig.~\ref{fig:lump_0_3} but for a different two dimensional subspace. 
}
\label{fig:lump_6_9}
\end{figure}

\begin{figure}
\includegraphics[scale=0.9]{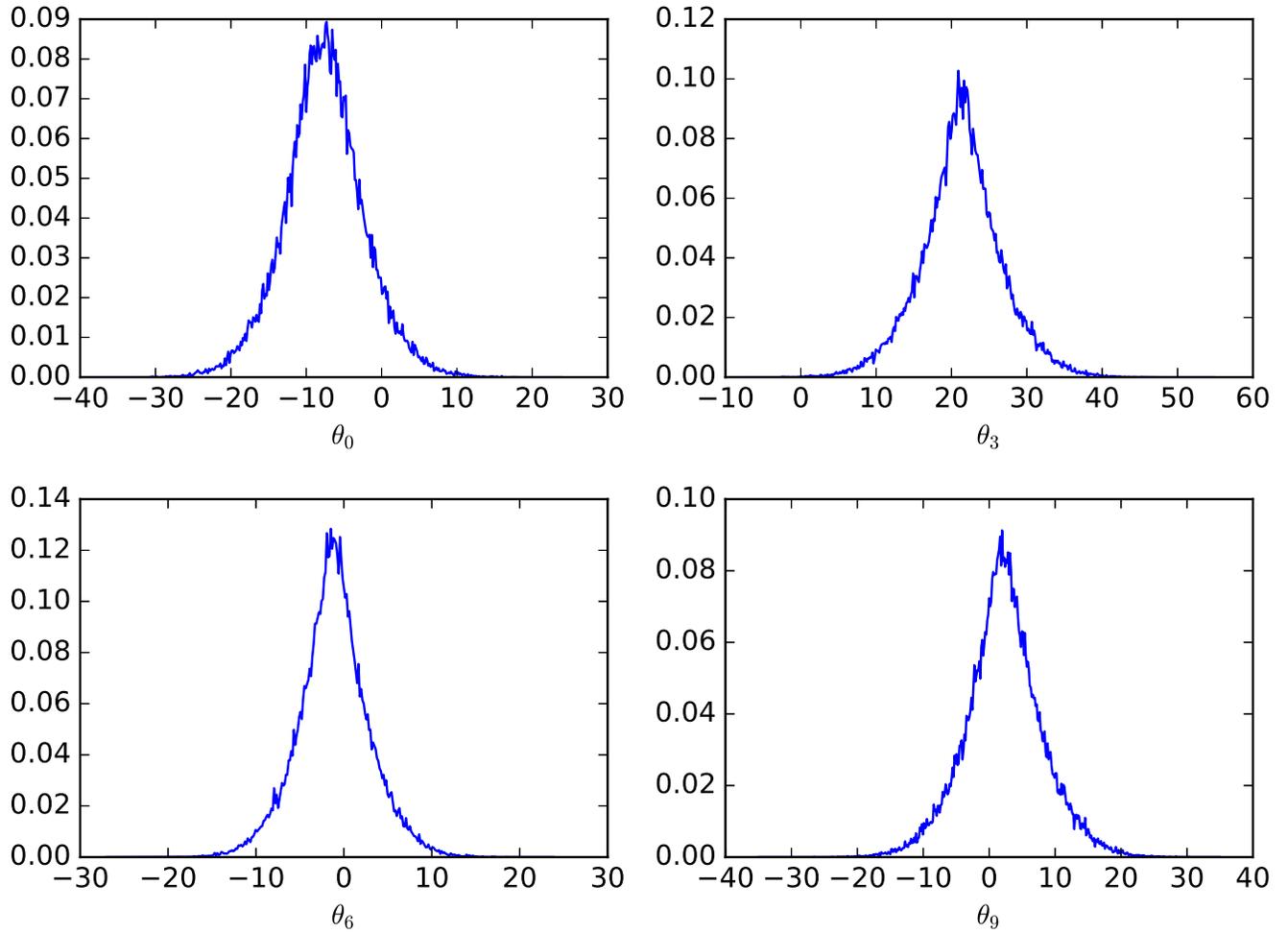}
\caption{
Four of the twelve marginalized one-dimensional posteriors
of the likelihood function plotted in Figures \ref{fig:lump_nlive},
\ref{fig:lump_0_3}, and \ref{fig:lump_6_9} as found by MultiNest.
Note the non-Gaussianity of the wings of the posteriors.
}
\label{fig:lump_1d}
\end{figure}

\begin{figure}
\includegraphics[scale=0.9]{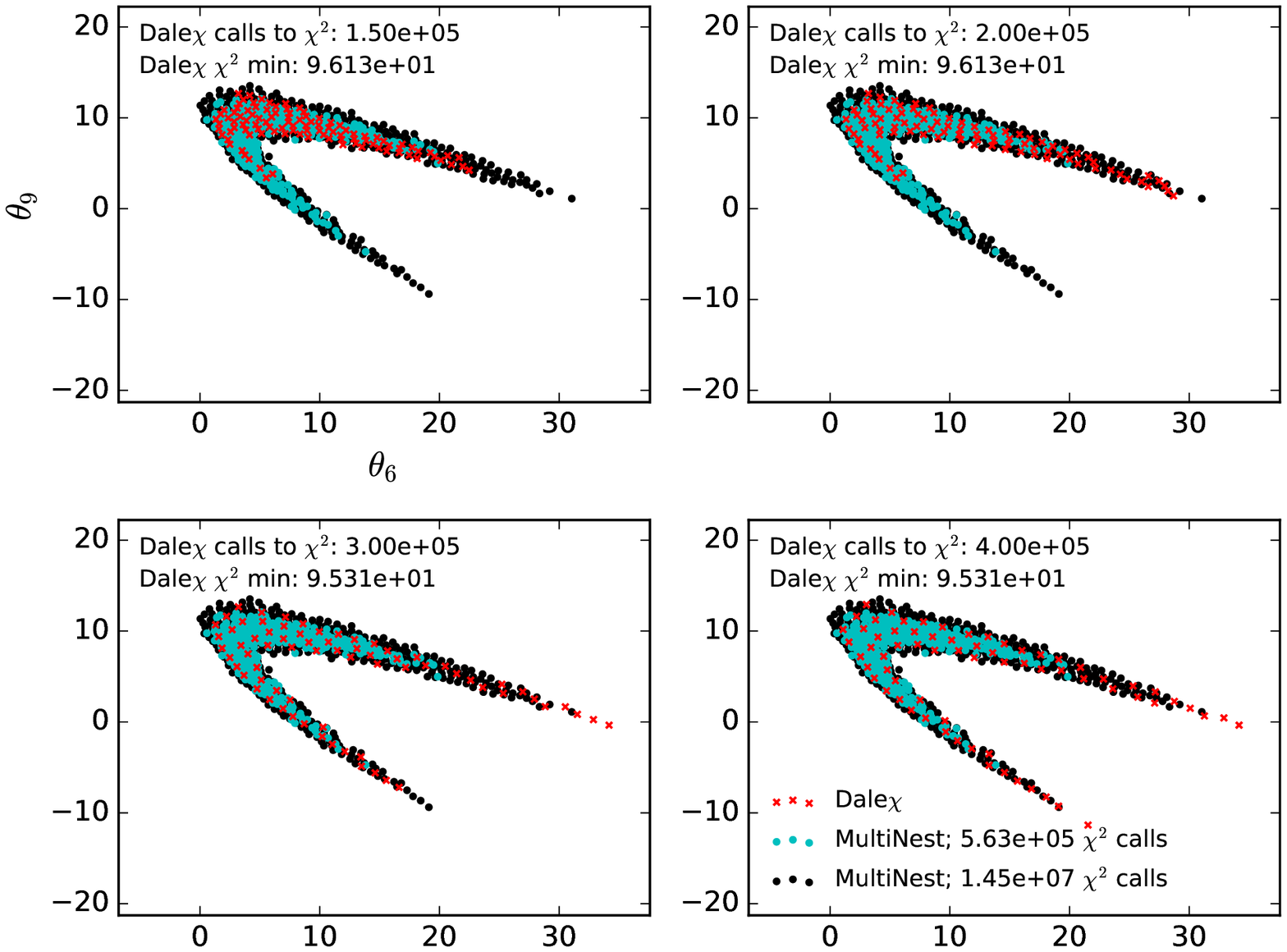}
\caption{
As Fig.~\ref{fig:lump_0_3} but for the curved degeneracy likelihood function. 
}
\label{fig:jellybean_0_1}
\end{figure}

\begin{figure}
\includegraphics[scale=0.9]{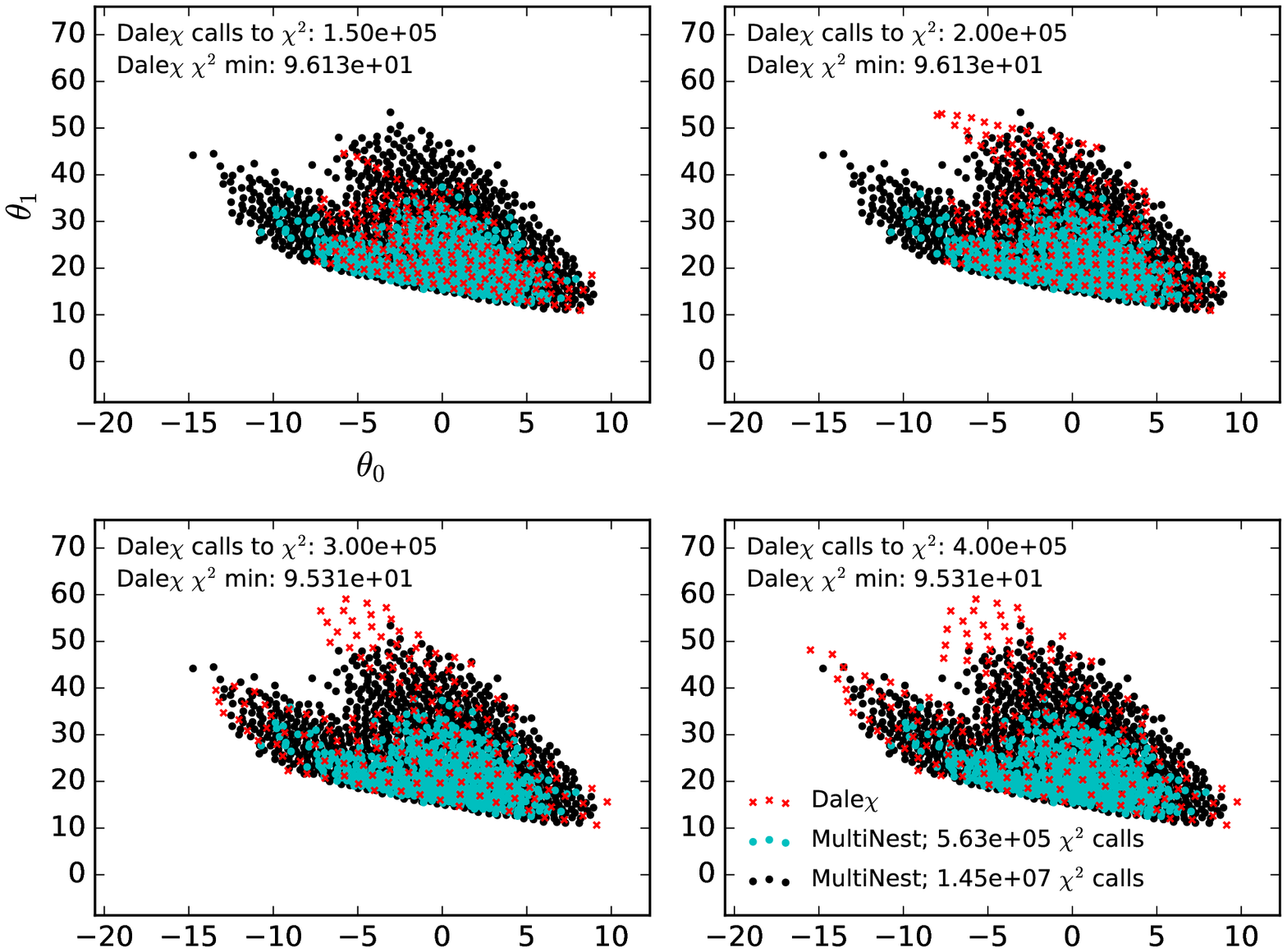}
\caption{
As Fig.~\ref{fig:jellybean_0_1} but for a different two dimensional subspace. 
}
\label{fig:jellybean_6_9}
\end{figure}

\begin{figure}
\includegraphics[scale=0.9]{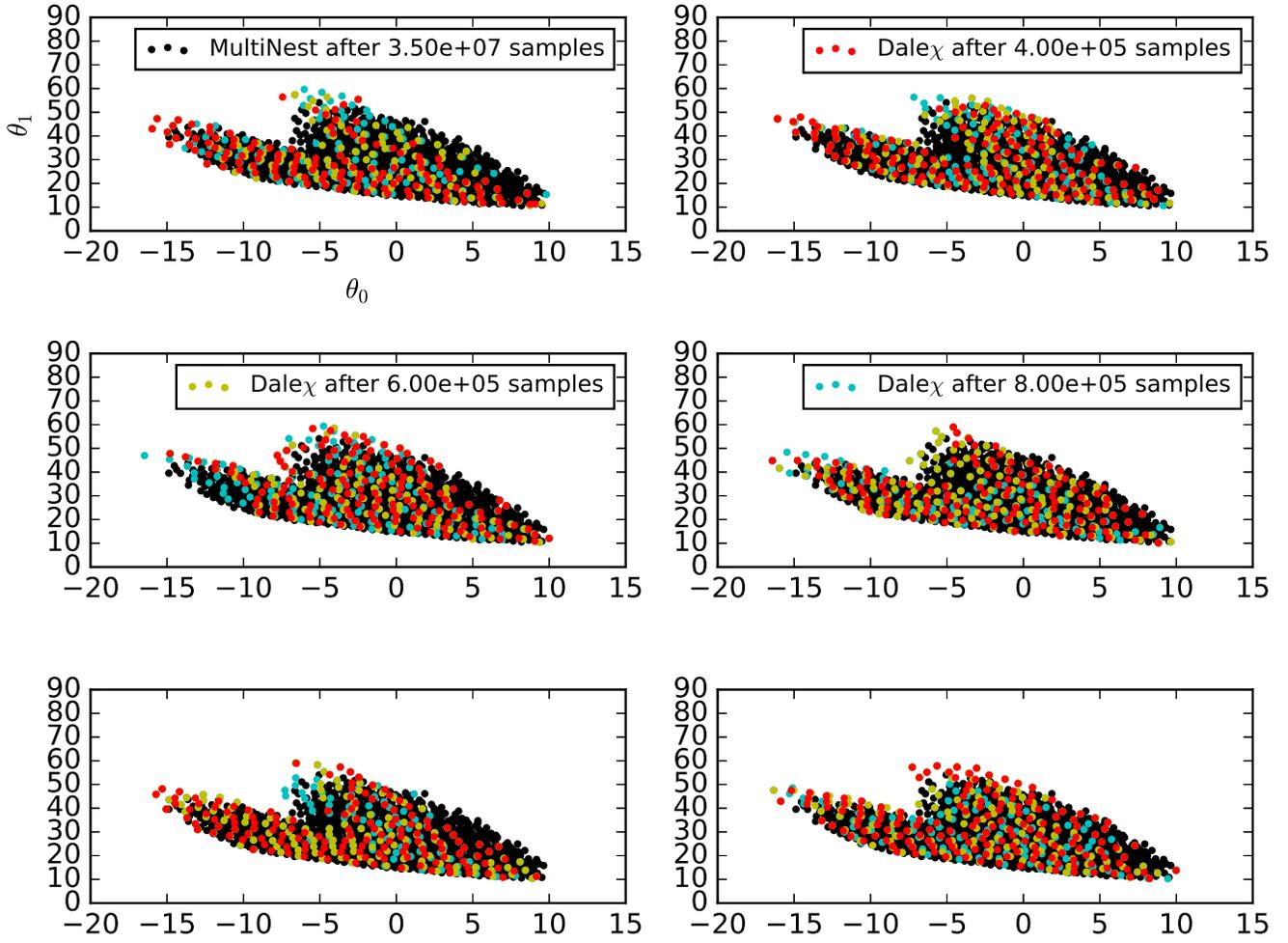}
\caption{
Each panel represents a different \ourCode run with a different
seed provided to the random number generator.  The colors show
the 95\% confidence limit as found at different stages during
the run's history.  The likelihood function is the same as
that tested in Figures \ref{fig:jellybean_0_1} and
\ref{fig:jellybean_6_9}. 
}
\label{fig:jellybean_evolution}
\end{figure}

Note that both MultiNest and \ourCode depend on random number generators: 
MultiNest for driving sampling of the posterior, \ourCode for driving
the Metropolis-Hastings-like algorithms described in Sections \ref{sec:optimize}
and \ref{sec:refine}, and for selecting seeds for the simplex minimizations
described in Sections~\ref{sec:explore} and \ref{sec:tendril}.  We have found
that the convergence performance of \ourCode can depend on how that random
number generator is seeded.  Figure \ref{fig:jellybean_evolution} plots
six different runs of \ourCode, each with a different random number
seed.  The runs plotted were specifically selected to represent instances
of \ourCode that did not find the full $(\theta_0,\theta_1)$ confidence
limit after 400,000 $\chi^2$ evaluations (all six successfully found the
$(\theta_6, \theta_9)$ confidence limit in 400,000 samples).  In each panel,
we show the confidence limit as found by \ourCode after 400,000, 600,000,
800,000 $\chi^2$ evaluations plotted against the true credible limit as
discovered by a run of MultiNest with 50,000 live points, taking 35 million 
evaluations.  In each case,
the 400,000 evaluation confidence limit traces out the general shape of the
true contour while the subsequent evaluations allow \ourCode to fill in the
gaps in that first discovered confidence limit.  Referring back to
Figure \ref{fig:jellybean_nlive}, we see that, even after 4.3 million $\chi^2$
evaluations, MultiNest has not found its final credible limit.  Therefore,
even these instances of sub-optimal runs of \ourCode demonstrate a significant speed-up
vis-\`a-vis MultiNest.  The difficulty is in knowing how many $\chi^2$
evaluations is enough for \ourCode.  This is the danger in not having
a clear global convergence criterion for \ourCode and a subject worthy
of further study.

For those interested, and in the interest of giving credit where credit
is due, \ourCode uses the random number generator proposed
by \citet{random} with initialization parameters proposed by \citet{nr}.
This is implemented in the class \texttt{Ran} defined in
\texttt{include/goto\_tools.h} in our code base.

\subsection{Efficiency of Sampling} \label{sec:sampling} 

\begin{figure}
\includegraphics[scale=0.9]{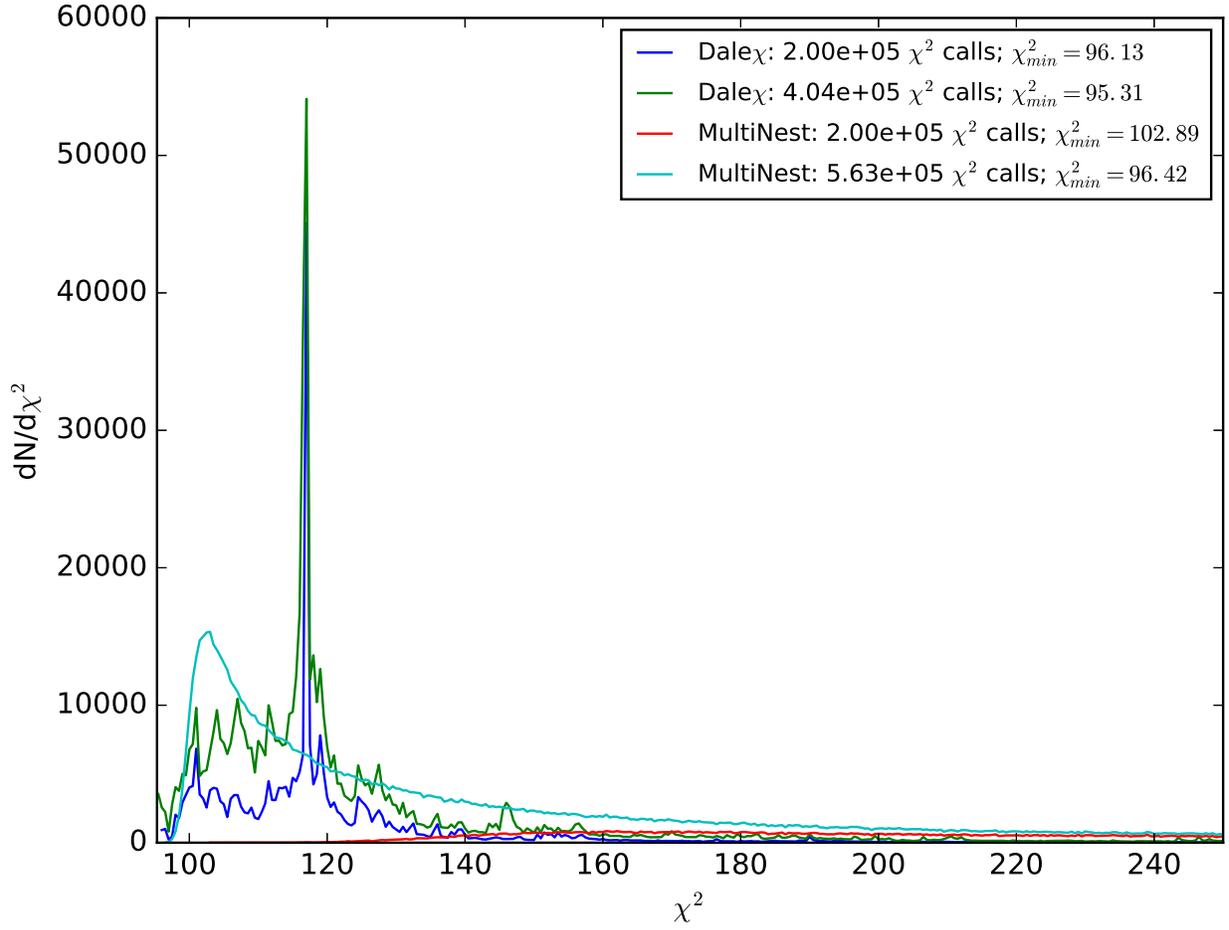}
\caption{
The distribution of points sampled by \ourCode and MultiNest
as a function of $\chi^2$ value at different points during
the runs' histories. 
}
\label{fig:histograms}
\end{figure}

Figure \ref{fig:histograms} shows the
distribution points evaluated by \ourCode and MultiNest as a function
of $\chi^2$ when run on the highly curved likelihood function from 
Figures~\ref{fig:jellybean_nlive},
\ref{fig:jellybean_0_1}, and \ref{fig:jellybean_6_9}.
Note how \ourCode places a much larger fraction
of its points on the $\chi^2=\clim$ contour than does MultiNest.
This should not be surprising as \ourCode was designed with limit-finding in
mind.  MultiNest is a code designed to calculate the integrated
Bayesian evidence of a model.  The fact that one can draw confidence limit
contours from the samples produced by MultiNest is basically a happy side-effect
of this evidence integration.  Because it focuses on finding the contours
directly, rather than integrating the posterior over parameter space,
\ourCode is highly efficient at both the optimization and 
characterization tasks, i.e.\ the key elements of determining the best 
fit and its error estimation, as designed. 
This tradeoff between obtaining the two critical elements for parameter 
estimation versus mapping the posterior more broadly (and sometimes less 
accurately) is the root cause of the vastly improved speed  
demonstrated by \ourCode.

\section{Conclusions} \label{sec:concl} 

As our exploration of the universe becomes increasingly detailed and 
precise, the range of astrophysical, instrumental, and theoretical 
systematics that must be taken into account require high dimensional 
parameter spaces. Accurate estimation of the desired parameters needs 
fast and robust techniques for exploring high dimensional, frequently 
nonlinearly degenerate and non-Gaussian joint probability distributions. 
\ourCode is designed with two paramount purposes: {\it optimization\/} -- 
finding the probability maximum, and {\it characterization\/} -- mapping 
out a specified confidence limit $\clim$. 

We have shown that its blend of strategies: to obtain robustly the 
maximum, search for multiple maxima, focus on the desired confidence 
limit, and efficiently map it out along multiple dimensions by seeking 
points distinct from established points and tracing along curved 
degeneracies -- can be quite successful. This diversity is 
important in cases where the likelihood is not known to be close to 
Gaussian but may have complicated, degenerate/highly curved/non-Gaussian, 
or multimodal properties. \ourCode exhibits excellent 
completeness and purity: it routinely finds viable regions of the 
posterior missed by Markov Chain Monte Carlo or MultiNest nested 
sampling, and it focuses its search in the critical areas rather than 
accreting points not directly relevant to the maximum or confidence 
limit. 

In the low, four dimensional test case, \ourCode had better purity 
(i.e.\ accuracy of confidence limit estimation) but MultiNest was 
faster. By the more common case of a 12 dimensional parameter space, 
\ourCode achieve a speed advantage of nearly a factor of 10. 
Note that \ourCode looks for a user specified confidence limit, e.g.\ 
95\%; if the user wants multiple confidence limits, e.g.\ also 68\%, 
then it is true that \ourCode must be rerun while MultiNest can generate 
it from its original results. However, the factor 10 speed advantage 
(coupled with that the first run of \ourCode will have already completed 
the optimization step of finding the probability maximum), means that 
even for calculation of several confidence limits the speed advantage 
remains. With higher dimensional parameter spaces we expect the advantage 
to become even stronger (unless the likelihood function is close to 
Gaussian in all dimensions).  A preliminary test on a 16-dimensional
likelihood function showed a factor of 75 speed advantage to \ourCode
over MultiNest. At a minimum, \ourCode can function as a highly
efficient ``pre-burner'' for sampling methods. 

Despite its current success, several areas remain open for further 
improvement to \ourCode. It seems likely that the $\cmin$ refinement 
procedure could be improved in efficiency.
The convergence criterion is local and somewhat 
heuristic; potentially a more robust condition could further improve 
performance. 
Note that \ourCode is open source so we encourage interested researchers 
to explore its use and join in to make it even better. 
It serves as a new tool in our quest to estimate accurately the parameters 
of our universe. 

The Direct Analysis of Limits via the Exterior of $\chi^2$ (\ourCode) 
has as its mission to look for the terminus of the desired confidence region, 
the boundary separating it from the unfavored exterior parameter space. 
One could say that in its quest to keep only the viable models, by terminating 
the exterior, the watchword for \ourCode is ex-terminate. Finally, 
\ourCode is available at \repo.

\acknowledgments 

We thank Alex Kim for helpful discussions. 
SD is supported by the LSST Coropration (\url{https://www.lsst.org}).
EL is supported in part by the Energetic Cosmos Laboratory and by 
the U.S.\ Department of Energy, Office of Science, Office of High Energy 
Physics, under Award DE-SC-0007867 and contract no.\ DE-AC02-05CH11231. 

\appendix 

\section{Fitting ellipsoids to samples} \label{sec:ellipsoid}

In order to prevent \ourCode from wasting time exploring regions of
$\chi^2\le\clim$ parameter space that have already been discovered,
we require a construct that represents the region in parameter space
bounded by a set of sampled points $\{\vec{P}\}$ in such a way that
we can quickly determine whether a new point $\vec{\theta}$ is inside
(in which case, \ourCode is doubling back on itself and must be
corrected) or outside (in which case, \ourCode has discovered a new
region of $\chi^2\le\clim$) that region.  The simplest way to represent
such a region would be to construct a hyperbox whose edges are at
the extremal values of each cardinal dimension as described by the
points in $\{\vec{P}\}$.  This, however, would involve erroneously
designating too much unexplored volume as ``explored''.  Consider
the case where the samples in $\{\vec{P}\}$ are tightly clustered around
the diagonal of a three-dimensional cube.  Using the extremal values
of $\{x, y, z\}$ in $\{\vec{P}\}$ would entail designating the entire
cube as ``explored'' when, in fact, we know almost nothing about the
cube as a whole.  We therefore have implemented the following algorithm
to take the set of sampled points $\{\vec{P}\}$ and determine a nearly
volumetrically minimal ellipsoid containing $\{\vec{P}\}$.  This is the
region that we designate as ``explored'' for the purposes of \ourCode.

The first step in designating our ellipsoid is to find the center of
the ellipsoid.  To do this, we find the extremal values of $\{\vec{P}\}$
in the cardinal dimensions ($\{x, y, z\}$ in a three-dimensional example).
We determine the geometric center of $\{\vec{P}\}$ to be at the point
\begin{equation}
\vec{\gamma}_i = 0.5\times\left(P_{i,\text{max}} + P_{i,\text{min}}\right) \ , 
\end{equation}
where $i$ is an index over the $D$ dimensions of parameter space.
We set the center of our ellipsoid to be the point in $\{\vec{P}\}$
that is closest to $\vec{\gamma}$ where, for the purposes of this
determination, we calculate the distance $\delta$ to be
\begin{equation}
\delta^2(\vec{\gamma},\vec{\theta}) =
\sum_i^D\left(\frac{\gamma_i-\theta_i}{P_{i,\text{max}}-P_{i,\text{min}}}\right)^2 \ . 
\end{equation}
Note the dimensions in parameter space are normalized by the denominator 
so each dimension runs from 0 to 1, putting them on an equal footing. 

Once we have found the center of our ellipsoid, which we call $\vec{c}$,
we must find the basis directions along which the axes of our ellipsoid will be
reckoned.  Initialize an empty set of $D$-dimensional vectors $\{\vec{B}\}$.
Iterate over the points $\vec{p}$ in $\{\vec{P}\}$.  For each point, calculate
the vector
\begin{equation}
\vec{\mu}=\vec{p}-\vec{c} -
\sum_{\vec{\beta}}\vec{\beta}\left[\left(\vec{p}-\vec{c}\right)\cdot\vec{\beta}\right]
\end{equation}
where $\vec{\beta}$ are the orthogonal unit vectors already in $\{\vec{B}\}$.
Select the vector $\vec{\mu}$ with the largest L2 norm.  Normalize
this vector and add it to $\{\vec{B}\}$.  Continue until there are $D$
vectors in $\{\vec{B}\}$.  In this way, we select a set of orthogonal
basis vectors along which the points $\{\vec{P}\}$ are extremally
distributed.

Now that we have determined the center and bases of the coordinate
system in which we will draw our ellipsoid, we must find a set of
lengths $\{r\}$ of the ellipsoid's axes such that all of the points in
$\{\vec{P}\}$ are contained in the ellipsoid.  The length of the
vectors $\vec{\mu}$ chosen to populate $\{\vec{P}\}$ provide a good
first guess at this set of lengths, but they are only guaranteed
to describe the ellipsoid if the points in $\{\vec{P}\}$ really are
only distributed along the vectors $\{\vec{B}\}$ and centered on
$\vec{c}$.  Since this seems unlikely, we adopt an iterative
scheme to adjust the lengths $\{r\}$ so that our ellipsoid
contains $\{\vec{P}\}$.  For each of the points $\vec{p}_\text{bad}$
that are not contained by the ellipsoid, find the basis vector
$\vec{\beta}_\text{worst}$ that maximizes
\begin{equation}
\epsilon = \frac{|\left(\vec{p}_{\text{bad}}-\vec{c}\right)\cdot\vec{\beta}|}{r_{\vec{\beta}}}
\end{equation}
where $r_{\vec{\beta}}$ is the length in $\{r\}$ corresponding to that basis
vector.  Keep track of the number of points $\vec{p}_\text{bad}$ for which
any given basis vector is $\vec{\beta}_\text{worst}$.  Select the basis
direction that is $\vec{\beta}_\text{worst}$ for the most points
$\vec{p}_\text{bad}$.  Multiply the corresponding $r$ by 1.1.
Continue this process until there are no more points $\vec{p}_\text{bad}$.

This algorithm is implemented in \texttt{include/ellipse.h} and
\texttt{src/utils/ellipse.cpp} in our code base.


\label{lastpage}


\begin{thebibliography}{99}

\bibitem[Bentley (1975)]{kdtree}
Bentley,~J.~L. 1975, Communications of the Association for Computing Machinery
{\bf 18}, 509

\bibitem[Brooks and Gelman (1998)]{BrooksGelman:1998}
Brooks, S. and Gelman A. 1998,
Journal of Computational and Graphical Statistics {\bf 7}, 434

\bibitem[Bryan (2007)]{brentsthesis}
Bryan, B., 2007, Ph.D. thesis, Carnegie Mellon University,
\url{http://reports-archive.adm.cs.cmu.edu/anon/ml2007/abstracts/07-122.html} 

\bibitem[Bryan {\it et al.\/} (2007)]{Bryan:2007}
Bryan, B., Schneider, J., Miller, C.~J., Nichol, R.~C., Genovese, C., and
Wasserman, L., 2007,
Astrophys. \ J. {\bf 665}, 25

\bibitem[Casella and George (1992)]{gibbs}
Casella, G. and George, E. I. 1992,
The American Statistician {\bf 46}, 167

\bibitem[Daniel {\it et al.\/} (2014)]{Daniel:2014}
Daniel, S.F., Connolly, A.J., and Schneider, J. 2014,
ApJ {\bf 794} 38 [arXiv:1205.2708]

\bibitem[Feroz and Hobson (2008)]{Feroz:2008}
Feroz, F. and Hobson, M.P. 2008,
Monthly Notices of the Royal Astronomical Society {\bf 384}, 449

\bibitem[Feroz {\it et al}. (2009)]{Feroz:2009}
Feroz, F., and Hobson, M. P., and Bridges, M. 2009,
Monthly Notices of the Royal Astronomical Society {\bf 398}, 1601

\bibitem[Feroz {\it et al}. (2013)]{13062144}
Feroz, F., Hobson, M. P., Cameron, E., and Pettitt, A. N.,
arXiv:1306.2144

\bibitem[Gelman and Rubin (1992)]{GelmanRubin:1992}
Gelman, A. and Rubin, D. 1992, Statistical Science {\bf 7}, 457

\bibitem[Kirkpatrick {\it et al}. (1983)]{annealing}
Kirkpatrick, S., Gelatt, C. D., and Vecchi, M. P. 1983,
Science {\bf 220}, 671

\bibitem[Marsaglia (2003)]{random}
Marsaglia,~G. 2003, Journal of Statistical Software, {\bf 8} 14

\bibitem[Nelder and Mead (1965)]{simplex}
Nelder,~J.~A. and Mead,~R. 1965, The Computer Journal, {\bf 7} 308

\bibitem[Planck Collaboration (2013)]{planck16} 
Planck Collaboration XVI 2014, Astronomy \& Astrophysics {\bf 571} 16 [arXiv:1303.5076] 

\bibitem[Press {\it et al}. (2007)]{nr}
Press,~W.~H., Teukolsky,~S.~A., Vetterling,~W.~T. and Flannery,~B.~P. 2007,
``Numerical Recipes'' (3d edition), Cambridge University Press, 2007 

\bibitem[Rasmussen and Williams (2006)]{gp}
Rasmussen, C.~E. and Williams, C.~K.~I., 2006, ``Processes for Machine Learning'' 
\url{http://www.GaussianProcess.org/gpml} 

\bibitem[Skilling (2004)]{skilling} 
Skilling J. 2004, in ``Bayesian Inference and Maximum Entropy Methods in
Science and Engineering'', eds. Fisches R., Preuss R., von Toussaint U., 
AIP Conf. Proc. 735, 395 

\bibitem[Wilks (1938)]{wilks}
Wilks,~S.~S. 1938, Annals of Mathematical Statistics, {\bf 9} 60

\end{thebibliography}
\end{document}